\documentclass[
  aps,
  prl,
  reprint,
  superscriptaddress,
  amsmath,
  amssymb,
  longbibliography
]{revtex4-2}

\usepackage{graphicx}
\usepackage{xcolor}
\usepackage{bm}
\usepackage{mathtools}
\usepackage{hyperref}
\usepackage{bibunits}
\usepackage{enumitem}
\usepackage{booktabs}
\usepackage{orcidlink}
\usepackage{caption}
\usepackage{comment}

\newcommand{\kmsMpc}{\mathrm{km\,s^{-1}\,Mpc^{-1}}}
\newcommand{\DGW}{\mathcal D_{\rm GW}}
\newcommand{\Dgal}{\mathcal D_{\rm g}}

\newcommand{\PRLparagraph}[1]{%
  \paragraph{\textbf{#1}\hspace{0.15em}---\hspace{0.15em}}%
}

\newcommand{\RESULTTODO}[1]{\textcolor{magenta}{\textbf{[RESULT: #1]}}}
\newcommand{\dd}{\mathrm d}

\makeatletter

\def\hyper@natlinkstart#1{%
  \Hy@backout{#1}%
  \hyper@linkstart{cite}{cite.\@bibunitname.#1}%
  \def\hyper@nat@current{#1}%
}

\def\hyper@natlinkbreak#1#2{%
  \hyper@linkend#1%
  \hyper@linkstart{cite}{cite.\@bibunitname.#2}%
}

\def\hyper@natanchorstart#1{%
  \Hy@raisedlink{%
    \hyper@anchorstart{cite.\@bibunitname.#1}%
  }%
}

\makeatother

\begin{document}

\newcommand{\MM}[1]{{\color{blue}MM: #1}}

\newcommand{\Ale}[1]{{\color{red}AA: #1}}

\begin{bibunit}[apsrev4-2]

\title{Gravitational-wave dark sirens as astrophysical probes: inferring galaxy–merger relations and identifying individual hosts}
%

\author{Alessandro Agapito \orcidlink{0009-0005-9004-3163}}
\author{Michele Mancarella \orcidlink{0000-0002-0675-508X}}

\affiliation{Aix-Marseille Universit\'e, Universit\'e de Toulon, CNRS, CPT, Marseille, France}

\begin{abstract}
%
%
%
Gravitational-wave (GW) ``dark siren'' techniques, namely the statistical
association of GW events with a galaxy catalog, have so far been used as
a cosmological tool.
Yet this association carries distinctive astrophysical information that
remains unused---specifically, how compact binaries populate galaxies and the
identities of individual hosts.
In this work, we turn the dark-siren program into an astrophysical probe by
inferring both simultaneously within a single hierarchical framework.
%
%
%
The method reduces the many potential hosts to a compact,
probabilistically ranked target list, providing a direct follow-up
decision tool.  
Applied blindly to GW170817, it recovers NGC~4993 as the highest-ranked host and constrains
the galaxy luminosity weighting, mildly favoring non-uniform host
weights.
Fixing the galaxy weighting can instead degrade host identification and even
favor the wrong host.
This framework recasts dark sirens as a potentially unique probe of the
astrophysics of electromagnetically dark compact-binary mergers.
%
\end{abstract}


\maketitle

\PRLparagraph{Introduction} Gravitational-wave (GW) dark-siren techniques statistically associate each GW event with galaxies compatible with its localization. The GW signal provides a luminosity-distance measurement, while a galaxy catalog supplies possible source redshifts ~\cite{Schutz:1986gp,DelPozzo:2011vcw}; candidate host galaxies are assigned relative probabilities according to their observed properties ~\cite{Chen:2017rfc,Gray:2019ksv,Finke:2021aom,Gair:2022zsa}. So far, this framework has been used to constrain cosmology
~\cite{Mastrogiovanni:2023emh,Gray:2023wgj,Borghi:2023opd, LIGOScientific:2025jau,LIGOScientific:2026uyd}, marginalising over the unknown host identity with a fixed assumption for the relative probability that a galaxy hosts a merger. 
However, both 
of these constitute a direct probe of compact-binary formation environments rather than nuisance parameters, as models of compact-binary formation predict correlations between merger rates and galaxy properties such as stellar mass, metallicity, and star-formation history~\cite{Artale:2020swx,Adhikari:2020wpn,Srinivasan:2023vaa}. 
This suggests reversing the usual logic and treating the GW--galaxy connection as an astrophysical signal in its own right rather than as a fixed input to cosmology. 

In this work, we turn this idea into a new dark-siren paradigm, developing a hierarchical framework that makes this astrophysical information directly and self--consistently measurable.
%
For sufficiently well-localized events, the resulting host probabilities can prioritize a small set of galaxies for targeted follow-up. Because an incorrect galaxy-host prescription can bias $H_0$~\cite{Hanselman:2024hqy,Perna:2024lod,Alfradique:2025tbj, Borghi:2025pav}, inferring such relation also self-calibrates a potential systematic in cosmological measurements.

A few studies have explored pieces of this broader possibility~\cite{Nuttall:2010nk,Fan:2014kka,Singer:2016erz,Vijaykumar:2023bgs,Hanselman:2024hqy,Perna:2024lod,Mo:2024uim,Li:2025hrh, Salvarese:2026usc}.
These approaches leave crucial parts of the GW--galaxy relation unused, either
%
not retaining event-level galaxy-catalog information, 
or assuming a given galaxy--merger relation. 
Only very recently a first framework to constrain galaxy weightings has been proposed~\cite{Gray:2026loe}.
To our knowledge, this is the first formulation of dark-siren inference that allows the galaxy--merger relation, individual host probabilities, and cosmology to be inferred jointly from the full GW and galaxy-catalog data within a single self-consistent hierarchy.

Accessing this full information opens a distinct, and for
electromagnetically dark binary black holes (BBHs) potentially unique, observational regime.
Electromagnetic counterparts can provide much stronger individual host
identifications but are rare and largely restricted to mergers
involving neutron stars, whereas GW--galaxy
cross-correlations~\cite{Raccanelli:2016cud,Scelfo:2018sny,Calore:2020bpd,
Libanore:2021jqv,Zazzera:2024agl,Pedrotti:2025tfg,Smith:2025bkq} probe
ensemble clustering rather than specific event--galaxy associations;
even with next-generation data, their uncertainties may remain too
large to distinguish confidently among different models of the
GW--galaxy connection~\cite{Pedrotti:2025tfg}.
Improving galaxy coverage and GW localization will substantially expand
the regime in which this inference is informative~\cite{LSSTScience:2009jmu,Euclid:2024yrr,WST:2024rai,
Borhanian:2020vyr,Branchesi:2023mws,Mo:2024frl,Dang:2025vqx,ET:2025xjr,
Punturo:2010zz,Reitze:2019iox}, making astrophysical use of dark
sirens increasingly accessible. 

\PRLparagraph{Full hierarchical GW--galaxy inference}

We use hierarchical Bayesian inference
~\cite{Mandel:2018mve,Vitale:2020aaz} to combine all GW events and the
galaxy catalog, consistently accounting for GW selection effects.
We denote the combined data by
$\mathcal D\equiv(\DGW,\Dgal)$, the source parameters of event $i$ ($i=1, \ \dots N_{\rm GW}$) by
$\theta_i$, and by $\lambda$ the $N_\lambda$ shared hyperparameters
describing the compact-binary population, cosmology, and galaxy--merger
connection.
A population term
$p_{\rm pop}(\{\theta_i\},\{g_i\},\{s_i\}\mid\lambda,\Dgal)$
specifies the joint distribution of the source and host variables for the ensemble of events, conditioned on the shared population hyperparameters and galaxy catalog.
This includes a prior on the source redshift and sky position, constructed by splitting
the redshift--sky distribution into catalogued and unresolved components, e.g.~\cite{Chen:2017rfc,Finke:2021aom}.
%
%

Crucially, the population prior retains for every event two discrete host variables.
The binary variable $s_i\in\{0,1\}$ specifies whether the true host is
represented in the galaxy catalog or belongs to the unresolved
population; conditional on $s_i=1$, $g_i$ identifies which catalog
galaxy hosts the event. We infer these variables jointly with the source and population parameters, so that catalog membership and individual host identity are themselves posterior quantities.
We adopt standard homogeneous completion~\cite{Finke:2021aom} for the
out-of-catalog component.

The merger--galaxy connection is encoded through a weight
$w(\mathcal G;\lambda_g)$ depending on observed galaxy properties
$\mathcal G$~\cite{Borghi:2025pav}. As a first realization, we take
$\mathcal G=L$, the galaxy luminosity derived from its observed apparent magnitude, 
and define
\begin{equation}
\label{eq:weight}
w(L,\epsilon)\propto
\left(\frac{L}{L_*}\right)^\epsilon ,
\end{equation}
where $L_*$ is a fixed pivot, taken to be the characteristic luminosity
of the 
Schechter function~\footnote{The interpretation of $\epsilon$ is conditional on the luminosity range
used to define the galaxy population: excluding galaxies below a fixed
threshold makes the inferred weighting a relative preference among the
galaxies above that threshold.}.
The same weight~\eqref{eq:weight} determines the relative
probabilities of catalog galaxies, the weighting of the unresolved
population, and their mixing fraction
$f_R(\epsilon;\Dgal)$~\cite{Chen:2017rfc,Finke:2021aom}.

We infer $\lambda_g=\epsilon$ as a
continuous population hyperparameter, thereby measuring the relative
preference of compact-binary mergers for galaxies of different
luminosities and enabling comparison with formation
models~\cite{Artale:2020swx,Adhikari:2020wpn,Srinivasan:2023vaa}.
%
%
%
%
%
Then, for each event, the catalog-membership variable $s_i$ 
is drawn with prior probability $p(s_i=1)=f_R(\epsilon;\Dgal)$. If $s_i=1$, $g_i$ then selects a catalog
galaxy and ties the source redshift and sky position to its measurements;
if $s_i=0$, these are drawn from the out-of-catalog distribution. 
%
%
%
%
The resulting posterior is a high-dimensional mixed discrete--continuous inference problem. If $N_\theta$ denotes the number of source parameters per event, the inference contains $(N_\theta+2)N_{\rm GW}+N_\lambda$ variables, where the two additional event-level variables are the host and catalog-membership labels.
Each $g_i$ is a single categorical variable, but it can take any
of $N_{\rm gal}^i$ candidate 
states; for example, across the simulated population used in this work, this amounts up to $\mathcal{O}(10^6)$ distinct host galaxies
explored jointly.
We implement the inference in \texttt{PyMC}~\cite{rainforth2017automating,pymc2023}, using a compound sampler that combines the No-U-Turn Sampler~\cite{JMLR:v15:hoffman14a} for the continuous variables with dedicated Gibbs-Metropolis~\cite{article} updates for the discrete variables. 
The implementation is available at~Ref.~\cite{pymcpop}. 
%
%
All details are described in the Supplemental
Material~\cite{SupplementalMaterial}

The physically relevant probability that
galaxy $g$ hosts event $i$ is its absolute posterior probability,
together with the complementary out-of-catalog probability,
\begin{equation}
\label{eq:pabs}
\begin{aligned}
p^{\rm abs}_{g,i}
\equiv &\ p(s_i=1,g_i=g\mid\mathcal D) = p^{\rm cond}_{g,i} \times p_{{\rm in}, i} \, ,\\
p^{\rm cond}_{g,i} \equiv &\ p(g_i=g\mid s_i=1,\mathcal D) \, , \\
p_{{\rm in}, i} \equiv &\ p(s_i=1\mid\mathcal D) \, , \\
p_{{\rm miss},i} \equiv &\ p(s_i=0\mid\mathcal D) = 1-p_{{\rm in},i}  \, , \\
\end{aligned}
\end{equation}
with $\sum_g p^{\rm abs}_{g,i}+p_{{\rm miss},i}=1$. The conditional probability $p^{\rm cond}_{g,i}$ ranks catalog galaxies under the hypothesis that the host is present. In the complete-catalog limit, $p_{{\rm miss},i}=0$ and
$p^{\rm abs}_{g,i}=p^{\rm cond}_{g,i}$.

Ranking galaxies by decreasing $p^{\rm abs}_{g,i}$ produces the
event-level observable relevant for follow-up: a probabilistically
ordered set of candidate hosts together with the probability that the
host is absent from the catalog. 
The ranking also determines the number of highest-ranked
galaxies required to enclose any chosen fraction of the host
posterior. All host probabilities are self--consistently marginalized over the inferred
galaxy--merger relation, compact-binary population, and cosmology.

\PRLparagraph{A+-era galaxy-weighting measurements and host identification}
We test the method on BBH populations embedded in the MICE mock galaxy
catalog~\cite{Carretero:2014ltj,Fosalba:2013wxa,Crocce:2013vda,
Fosalba:2013mra}, with injected $\epsilon_{\rm true}=0$ and $1$.
For each population we analyze 70 events detected by a
LIGO--Virgo--KAGRA--LIGO India (LVKI) network
~\cite{LIGOScientific2014pky,LIGOI,VIRGO2014yos,Aso2013eba} at design
A+ sensitivity, requiring network ${\rm S/N}>25$ and
$\Delta\Omega_{90}<2\,\mathrm{deg}^2$. This deliberately selects the
well-localized tail, where individual-galaxy inference is informative
and explicit host sampling remains computationally tractable; the same selection is included in the selection-function~\footnote{These samples are designed to validate this regime and should not be
interpreted as rate forecasts for a particular observing run.}.
We consider both the complete catalog and an incomplete
catalog obtained with an apparent-magnitude threshold
$m_{\rm th}=21$~\footnote{This threshold is not intended to reproduce a
specific survey. It produces an intermediate-completeness regime in
which both the observed- and missing-galaxy branches carry appreciable
posterior support while retaining sufficient structure for event-level
host inference.}, assuming spectroscopic redshift measurements.
We model GW detectability and parameter estimation with the
Fisher--matrix pipeline \texttt{GWFAST}
~\citep{Iacovelli:2022bbs,Iacovelli:2022mbg}. 
$H_0$ is the only varying
cosmological parameter.
Details 
are given in the Supplemental
Material~\cite{SupplementalMaterial}.

We first describe the main event-level product of the framework.
Figure~\ref{fig:population_cd} introduces an
operational follow-up decision tool plot that can be applied directly to
real GW events. 
For each event, $N_{50}^{\rm cat,i}$ and
$N_{90}^{\rm cat,i}$ give the number of highest-ranked galaxies
containing $50\%$ and $90\%$ of the posterior,  
while $p_{{\rm in},i}$ gives the probability that the
host is represented in the catalog.
Together, these quantities determine not only \emph{which} galaxies to
observe, but whether galaxy targeting is the appropriate next step.
Small $N_{50}^{\rm cat}$ and high $p_{\rm in}$ define the
{\sc target now} regime---few galaxies concentrate most of the likely catalogued-host probability and are thus worth immediate follow-up;
low $p_{\rm in}$ instead motivates completing the galaxy coverage first
({\sc complete first}). Large $N_{50}^{\rm cat}$ and high $p_{\rm in}$
define the {\sc prioritize} regime: the host is likely already
catalogued, but additional discriminating information---for instance
further galaxy properties or complementary observations---is advisable to
reduce the candidate set before targeted follow-up. If $N_{50}^{\rm cat}$ is large and $p_{\rm in}$ is low, deeper galaxy
coverage is instead the priority ({\sc deeper coverage}).
Notably, for a non-negligible fraction of events in our simulations, $90\%$ of the host posterior is concentrated in only $\mathcal O(10)$ galaxies while
$p_{\rm in}>90\%$, yielding compact and reliable target lists.

\begin{figure}
\includegraphics[width=.5\textwidth]{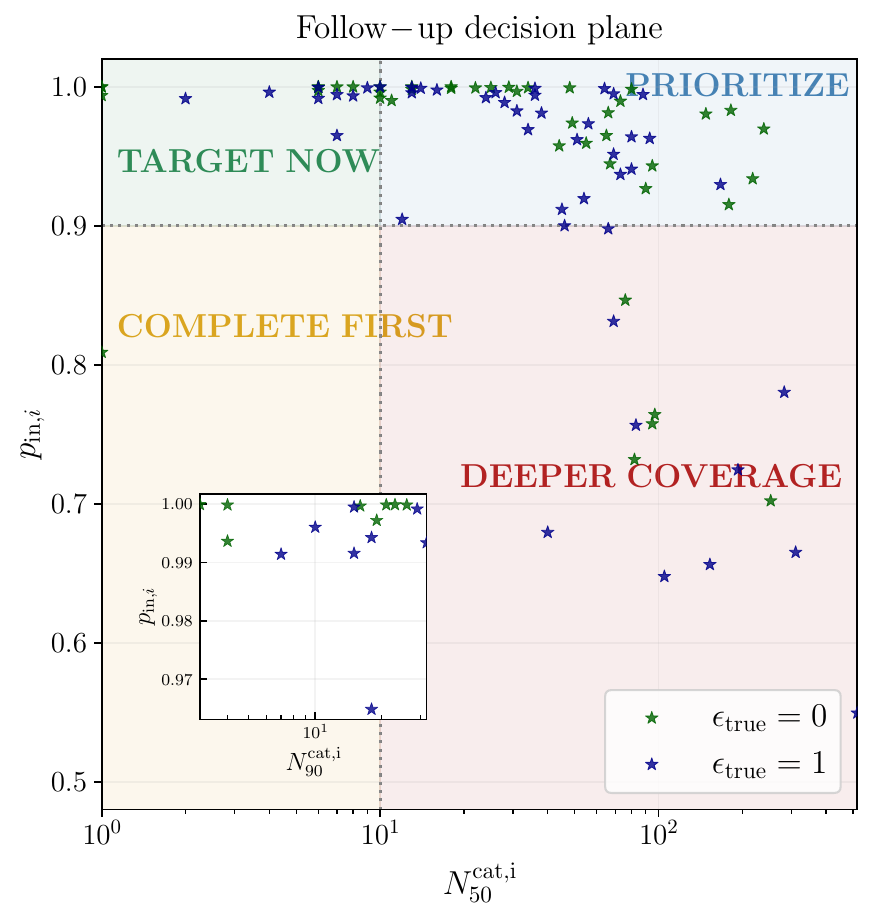}
\caption{Operational follow-up decision plane showing when a GW event supports
immediate galaxy follow-up and when improved coverage is needed in terms of both $N_{50}^{\rm cat}$ and $p_{\rm in}$.
Dotted lines separate the four follow-up regimes.
The inset shows $N_{90}^{\rm cat}$ for events in the {\sc target now}
region.
The regime boundaries are indicative and should be chosen
according to the observational goals and available resources.
}
\label{fig:population_cd}
\end{figure}
\begin{figure}
\includegraphics[width=.23\textwidth]{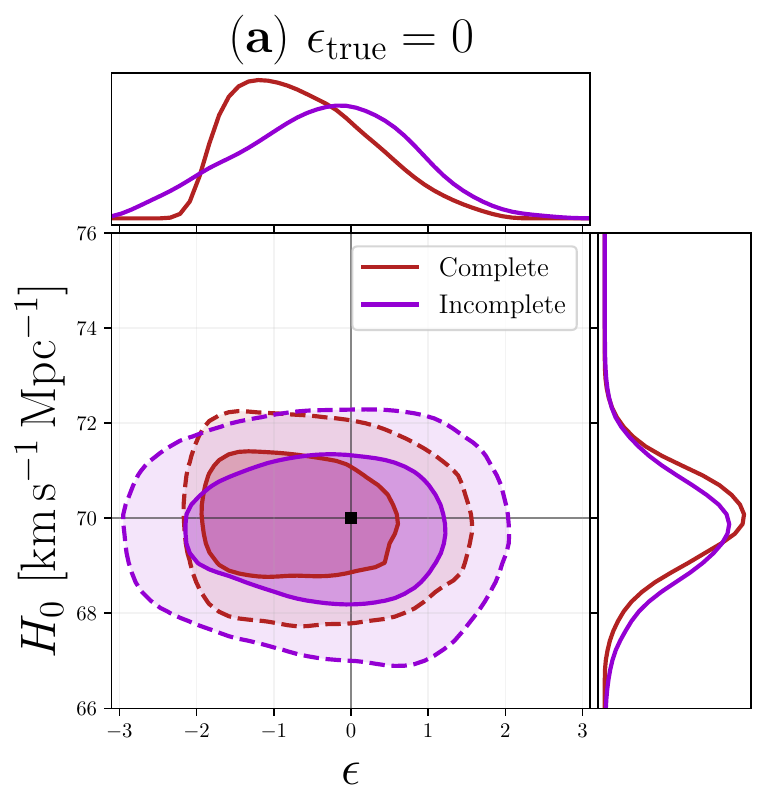}
\includegraphics[width=.23\textwidth]{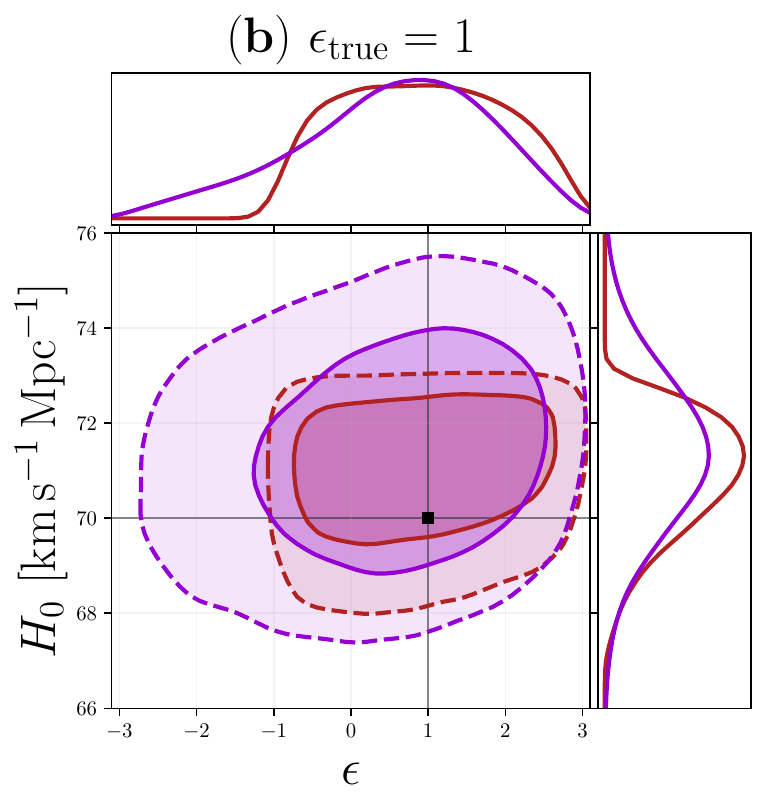}
\includegraphics[width=.46\textwidth]{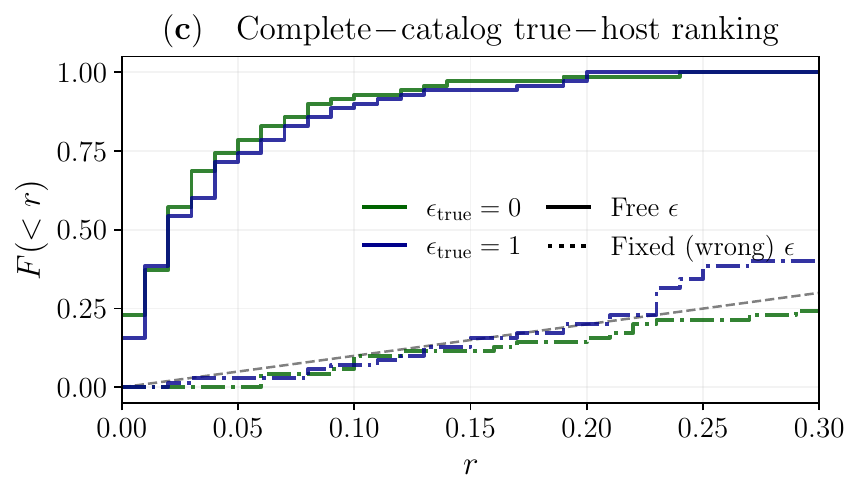}
\caption{Population- and event-level validation of the hierarchical GW--galaxy
inference on simulations.
(a,b) Joint $H_0$--$\epsilon$ posteriors for populations generated with
$\epsilon_{\rm true}=0$ and $1$, respectively, for complete and
incomplete galaxy catalogs; injected values are marked by the black
squares and reference lines.
Solid (dashed) lines represent $68\%$ ($95\%$) contours.
(c) Cumulative fractional rank of the true host for complete-catalog
events, comparing inference with free $\epsilon$ to a misspecified fixed
weighting; the diagonal corresponds to random association.}
\label{fig:population_ab}
\end{figure}

We also validate the underlying population- and event-level inference against the known simulated truth.
Figure~\ref{fig:population_ab} shows joint contours in the $\epsilon-H_0$ plane in panels (a)-(b).
The unbiased recovery of $\epsilon$
proves that the galaxy weighting can itself be measured from dark-siren
data, turning it into an astrophysical observable of the GW--galaxy
connection. 
Inferring $\epsilon$ jointly with $H_0$ also self-calibrates the
galaxy-weight prescription entering cosmological inference.

Finally, panel~(c) in Fig.\ref{fig:population_ab} quantifies the host-identification power of the
inferred probabilities through the cumulative fractional rank
$F(<r)$, where $r$ is the fraction of candidate galaxies ranked ahead
of the true host.
%
%
A curve above the diagonal indicates
better-than-random host association, with $r=0$ corresponding to the
true host being ranked first.
The true galaxy is ranked substantially better than random. For $\epsilon_{\rm true}=0$, it is ranked first in $23\%$ of events, lies within the top $5\%$ of candidates in $79\%$ of events, and within the top $10\%$ in $93\%$ of events; for $\epsilon_{\rm true}=1$, the corresponding fractions are $16\%$, $74\%$, and $90\%$.

Crucially, we find that incorrect assumptions on luminosity weighting impact the event-level host ranking. As shown in Fig.\ref{fig:population_ab} (c), fixing a misspecified $\epsilon$ drives the
ranking close to random for both injected populations.
The fraction of events in which the true host lies
within the top $10\%$ of candidates drops to $\sim 10\%$ for both $\epsilon_{\rm true}=0$ and 1.
when the weighting is misspecified. 
Further validation, including incomplete catalogs, is given in the
Supplemental Material~\cite{SupplementalMaterial}.


\begin{figure*}[t]
\centering
\includegraphics[width=1.\textwidth]{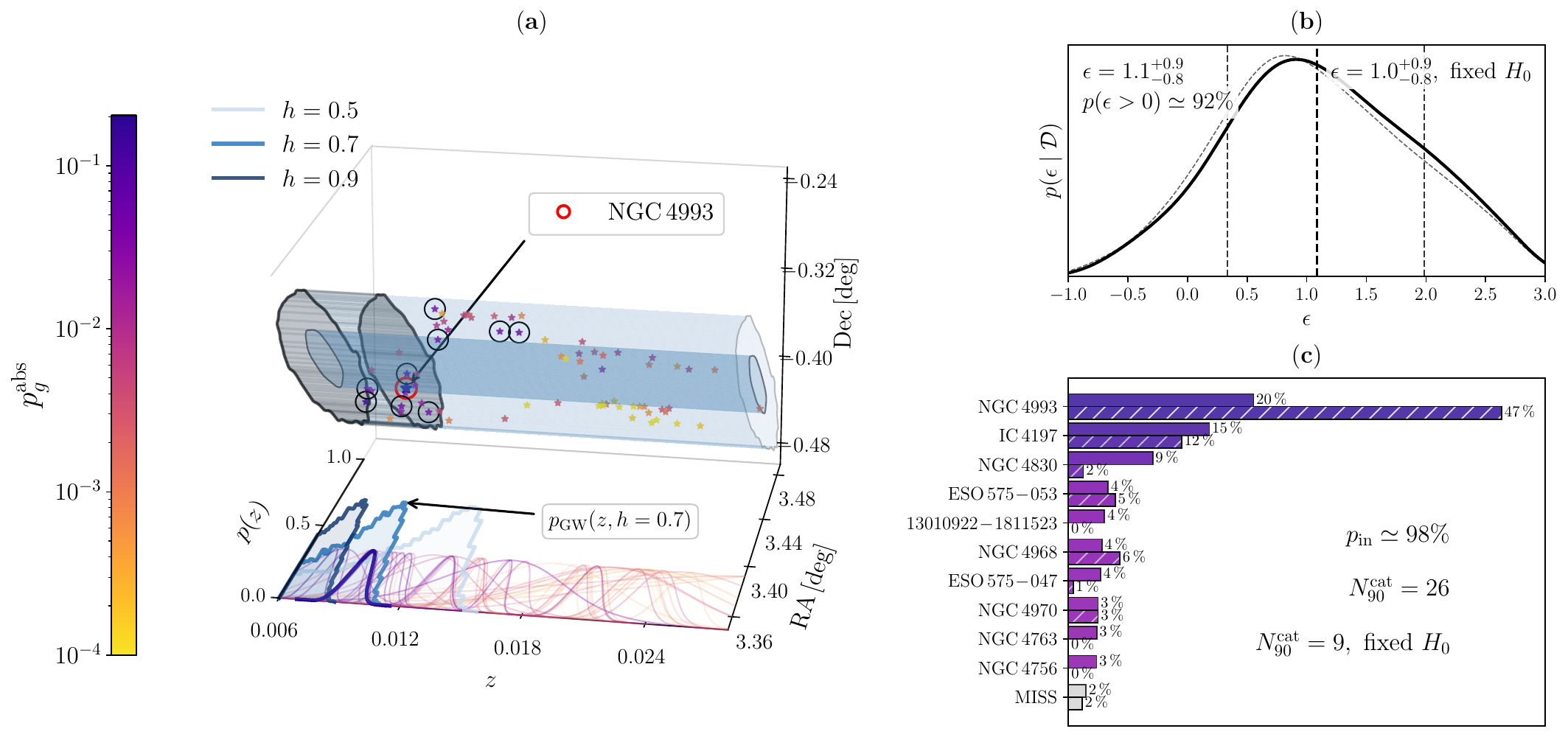}
\caption{
GW170817 treated as a dark siren.
(a) Candidate GLADE+ galaxies in sky position and redshift, colored by their absolute host probability $p_g^{\rm abs}$. NGC~4993 is marked by a red circle, while the other nine highest-ranked galaxies are marked by black circles. The light-blue tube shows the 50\% and 95\% credible GW volumes 
; the gray-shaded tube delimit the volume obtained assuming $H_0$ fixed to its Planck value. Projected redshift distributions show the GW redshift-mapped luminosity-distance posterior for representative values of $h = H_0/100\,\kmsMpc$ together with the galaxy redshift PDFs; NGC~4993 is highlighted. (b) Marginalized posterior for the luminosity-weighting exponent $\epsilon$ with median, 68\% credible interval, and $p(\epsilon>0)$ indicated. The dashed posterior correspond to the fixed-$H_0$ case. (c) Absolute posterior probabilities of the 9 highest-ranked catalog
galaxies and of the out-of-catalog hypothesis, ${\rm MISS}$; galaxy
labels give the corresponding follow-up targets. The lower, hatched bars correspond to the analysis with fixed $H_0$.
}
\label{fig:GW170817}
\end{figure*}
\PRLparagraph{GW170817 as a dark-siren}
We finally treat GW170817 as a dark
siren~\citep{LIGOScientific:2017vwq,LIGOScientific:2018gmd}, using the GW posterior including sky-localization information
without fixing the source position to the electromagnetic counterpart.
We construct the galaxy prior from $K$-band galaxies in
GLADE+~\cite{Dalya:2021ewn}, so that the independently known host
NGC~4993 provides a blind validation of the event-level inference.
We assume flat $\Lambda$CDM, varying $H_0$ while
fixing $\Omega_m$ to the Planck value~\cite{Planck:2018vyg}.
Details 
and robustness tests are given in the Supplemental
Material~\cite{SupplementalMaterial}.

Figure~\ref{fig:GW170817} summarizes the results. Panel~(a) makes the
physical origin of the host ranking explicit: consistency with the GW localization identifies plausible galaxies, while the inferred galaxy
weighting modulates how posterior support for the most likely host is distributed among them.
Panels~(b) and~(c) show,
respectively, the constraint on the luminosity-weighting exponent
$\epsilon$ and the top ranked absolute host probabilities, including the
out-of-catalog hypothesis.


Marginalizing over population and cosmology, we obtain
\begin{equation}
\epsilon=1.1^{+0.9}_{-0.8},
\qquad
p(\epsilon>0)\simeq0.92,
\end{equation}
where the interval denotes the central $68\%$ credible range. The
constraint is broad and not yet statistically compelling, but mildly
favors a luminosity-dependent over a uniform host weighting. 
Together with the recent BBH population constraint of
Ref.~\cite{Gray:2026loe}, this provides the first direct measurement of the galaxy--merger weighting for a BNS source, demonstrating that this astrophysical connection is already accessible to dark-siren inference.

At the event level, NGC~4993 is correctly ranked first among $N_{\rm gal}=236$ candidate catalog galaxies without imposing
its electromagnetic identification. For the fiducial luminosity cut, its conditional host probability is $p^{\rm cond}_{\rm NGC\,4993}\simeq 21\%$, while $p_{\rm in}\simeq 98\%$, 
and $p^{\rm abs}_{\rm NGC\,4993}\simeq 20\%$. 

We repeat the same analysis fixing $H_0$ to the Planck
value~\cite{Planck:2018vyg} to assess the added value of external
cosmological information. The absolute probability of NGC~4993
increases markedly to approximately $47\%$, widening the gap with the
second-ranked candidate by $27\%$, while the posterior on $\epsilon$
remains essentially unchanged. Most importantly for follow-up,
$N_{90}^{\rm cat}$ decreases to only 9 galaxies---just under
$\sim4\%$ of the candidate set---which together contain $90\%$ of the total host posterior.
The corresponding probabilities are displayed with hatched bars in Fig.~\ref{fig:GW170817} (c).


Finally, we point out the risks associated to a search for candidate hosts at fixed luminosity weighting.
The inferred host ranking is critically sensitive to both the selection threshold on the galaxy catalog and the assumed luminosity weighting.
We find that an overly aggressive luminosity cut ($L>1.5L_*$) favors the more luminous
NGC~4548 over NGC~4993; uniform weighting ($\epsilon=0$) favors
ESO575-053, while an overly aggressive weighting ($\epsilon=3$) instead favors NGC~4830. Thus all three misspecified choices produce a
confidently wrong highest-ranked host. Only moderate selections with
the intermediate $\epsilon=1$ correctly recover NGC~4993.
Notably, marginalizing over $\epsilon$ does not lead to statistically significant changes in the recovered $H_0$ relative to the fixed-weighting cases, with the resulting posterior remaining consistent with previous dark-siren analyses~\citep{LIGOScientific:2018gmd}.
A detailed discussion is given in the Supplemental
Material~\cite{SupplementalMaterial}.



\PRLparagraph{Perspective}
%

This work shows that GW dark sirens contain an astrophysical information
channel beyond cosmology that is unused but already exploitable. The framework
introduced here extracts the relevant astrophysical observables,
revealing how mergers populate galaxies and, for well-localized events,
reducing the host candidates to a probabilistically ranked target list
for follow-up.
The proof-of-principle analysis of this work points directly to a broader research program:

\begin{itemize}

\item \emph{The well-localized real-data frontier.}
Individual-host inference and population-level measurements need not be
optimized by the same events. Future work should map the trade-off
between localization, information on the galaxy--merger relation, and
computational cost. Applying the framework to the best-localized BBHs
in current and near-future LVK data is a concrete possibility,
as events localized to only a few square degrees are beginning to
emerge~\cite{LIGOScientific:2026ctl}. Such events could also motivate
dedicated galaxy observations of the localization region, while deeper
surveys and eventually third-generation GW networks will greatly expand
this regime. Our follow-up decision plane in Fig.~\ref{fig:population_cd} is the tool to guide such decisions.

\item \emph{Galaxy populations beyond luminosity.}
The present $w\propto L^\epsilon$ model is deliberately minimal.
Extending $w(\mathcal G;\lambda_g)$ to stellar mass, star-formation
rate, metallicity, color, morphology, or their combinations would
enable direct tests of galaxy-aware compact-binary formation models
and of whether different GW subpopulations occupy different galactic
environments. More generally, the hierarchy can describe heterogeneous
galaxy subpopulations rather than a single universal host relation.

\item \emph{Incompleteness and redshift uncertainties.}
Our treatment assumes a fixed non-evolving Schechter luminosity function and low-luminosity boundary. Both are coupled to the inferred
galaxy--merger relation: at the faint end, changes in the Schechter slope can mimic changes in the luminosity-weighting exponent, while the luminosity cut defines the galaxy population relative to which $\epsilon$ is measured. Future applications should
therefore infer these quantities jointly. Our proof-of-principle
implementation also assumes negligible angular variations in catalog
completeness; these will be incorporated in follw-up works. 
Finally, the present analyses use spectroscopic galaxy redshifts with
Gaussian likelihoods. Broader and non-Gaussian redshift posteriors can
be incorporated naturally within the same sampling scheme, which is
particularly relevant for photometric-redshift catalogs. For
well-localized events, an intermediate strategy could use such catalogs
for the initial inference and let the resulting host posterior identify
the most promising galaxies for targeted spectroscopic follow-up.

\end{itemize}

More broadly, this work opens a substantially wider observational role for dark sirens, extending a cosmological technique into a uniquely direct
probe of gravitational-wave astrophysics.


\vspace{1cm}

\begin{acknowledgments}

\paragraph{Acknowledgments.---}
We thank Sayantani Bera, Nicola Borghi, Maria Lisa Brozzetti, Sarah Ferraiuolo, Simone Mastrogiovanni for discussions, and Matteo Tagliazucchi for internal LVK review and useful comments. This work is supported by the French government under the France 2030 investment plan, as part of the Initiative d'Excellence d'Aix-Marseille Universit\'e--A*MIDEX AMX-22-CEI-02.
This document has LIGO DCC document number LIGO-XXXX.


{\small This research has made use of data or software obtained from the Gravitational Wave Open Science Center (gwosc.org), a service of the LIGO Scientific Collaboration, the Virgo Collaboration, and KAGRA. This material is based upon work supported by NSF's LIGO Laboratory which is a major facility fully funded by the National Science Foundation, as well as the Science and Technology Facilities Council (STFC) of the United Kingdom, the Max-Planck-Society (MPS), and the State of Niedersachsen/Germany for support of the construction of Advanced LIGO and construction and operation of the GEO600 detector. Additional support for Advanced LIGO was provided by the Australian Research Council. Virgo is funded, through the European Gravitational Observatory (EGO), by the French Centre National de Recherche Scientifique (CNRS), the Italian Istituto Nazionale di Fisica Nucleare (INFN) and the Dutch Nikhef, with contributions by institutions from Belgium, Germany, Greece, Hungary, Ireland, Japan, Monaco, Poland, Portugal, Spain. KAGRA is supported by Ministry of Education, Culture, Sports, Science and Technology (MEXT), Japan Society for the Promotion of Science (JSPS) in Japan; National Research Foundation (NRF) and Ministry of Science and ICT (MSIT) in Korea; Academia Sinica (AS) and National Science and Technology Council (NSTC) in Taiwan.
}

\end{acknowledgments}


\putbib[references,\jobname Notes]

\end{bibunit}


\clearpage
\onecolumngrid

\begin{bibunit}[apsrev4-2]

\setcounter{section}{0}
\setcounter{equation}{0}
\setcounter{figure}{0}
\setcounter{table}{0}

\renewcommand{\thesection}{S\arabic{section}}
\renewcommand{\theequation}{S\arabic{equation}}
\renewcommand{\thefigure}{S\arabic{figure}}
\renewcommand{\thetable}{S\arabic{table}}




\renewcommand{\theequation}{S\arabic{equation}}
\renewcommand{\thefigure}{S\arabic{figure}}
\renewcommand{\thetable}{S\arabic{table}}
\renewcommand{\thesection}{S\arabic{section}}
\renewcommand{\thesubsection}{\thesection.\arabic{subsection}}
\setcounter{secnumdepth}{3}




\title{Supplemental Material}
\maketitle
\onecolumngrid

This Supplemental Material provides the complete construction and numerical implementation of the hierarchical GW--galaxy model used in the main text. Section~\ref{secS:hierarchy} derives the galaxy-informed population distribution, including catalog incompleteness. Section~\ref{secS:sampling} gives the full-posterior sampling scheme and proposal corrections. Section~\ref{secS:sims} specifies the simulated galaxies, GW populations, and selection calculation. Sections~\ref{secS:popvalidation} and~\ref{sec:hostvalidation} collect population-level validation and  host-identification diagnostics. Section~\ref{secS:gw170817} gives the full GW170817 setup and related robustness tests.
Finally, section~\ref{secS:numerics} describes complementary convergence checks. 
\section{Full GW--galaxy hierarchical model and catalog incompleteness}
\label{secS:hierarchy}

We denote the GW data by $\DGW\equiv\{d_i\}_{i=1}^{N_{\rm ev}}$ and the galaxy-catalog data by $\Dgal\equiv\{x_g\}_{g=1}^{N_{\rm GAL}}$, where $x_g=(\hat z_g,\hat\Omega_g,\hat L_g)$ collects the measured redshift, sky position, and luminosity of catalog galaxy $g$. For event $i$, the source parameters are $\theta_i=(\bar\theta_i,z_i,\Omega_i)$, where $z_i$ is the source redshift, $\Omega_i$ the sky position, and $\bar\theta_i$ contains the remaining compact-binary parameters. Population and cosmological hyperparameters are collectively denoted by $\lambda$; the subset $\lambda_g$ describes the relation between mergers and galaxy properties. The population distribution entering the hierarchical posterior contains a galaxy-informed redshift--sky factor,
\begin{equation}
\label{eqS:popfactor}
p_{\rm pop}(\theta\mid\lambda,\Dgal)
\propto
p_{\rm CBC}(\bar\theta\mid\lambda)\,
\frac{\psi(z\mid\lambda)}{1+z}\,
p_{\rm gal}(z,\Omega\mid\lambda,\Dgal),
\end{equation}
where $p_{\rm CBC}(\bar\theta\mid\lambda)$ is the normalized distribution of the compact-binary parameters other than redshift and sky position, $\psi(z\mid\lambda)$ describes the source-frame merger-rate evolution, and $1/(1+z)$ converts to detector-frame time. The proportionality constant is fixed by normalizing $p_{\rm pop}$ over the population domain. In the following, we provide a full derivation of the galaxy catalog redshift prior $p_{\rm gal}$. 

\subsection{Redshift prior associated with galaxies}

Let $\mathcal G$ denote any set of galaxy properties that can affect both their intrinsic abundance and the probability of hosting a merger, such as luminosity (or magnitude), color, star formation rate, metallicity, or stellar mass~\citep{Borghi:2025pav}. The total number of mergers per unit redshift, solid angle, and galaxy properties can be written as~\cite{Mastrogiovanni:2023emh}
\begin{equation}
\label{eqS:CBC}
\frac{\dd N_{\rm CBC}}{\dd z\,\dd\Omega\,\dd\mathcal G}
=
T_{\rm obs}R_{0,*}
\frac{\psi(z\mid\lambda)}{1+z}
\,w(\mathcal G;\lambda_g)\,
\frac{\dd N_{\rm GAL}}{\dd z\,\dd\Omega\,\dd\mathcal G}.
\end{equation}
Here $T_{\rm obs}$ is the observing time, $R_{0,*}$ is an overall rate normalization per galaxy, and $w(\mathcal G;\lambda_g)$ is a relative host weight. Only relative weights matter for the normalized population distribution; the overall rate is marginalized in the hierarchical analysis.

After marginalizing over galaxy properties,
\begin{equation}
\label{eqS:effCBC}
\frac{\dd N_{\rm CBC}}{\dd z\,\dd\Omega}
=
T_{\rm obs}R_{0,*}
\frac{\psi(z\mid\lambda)}{1+z}
\int \dd\mathcal G\,
w(\mathcal G;\lambda_g)
\frac{\dd N_{\rm GAL}}{\dd z\,\dd\Omega\,\dd\mathcal G}.
\end{equation}
We decompose the galaxy population into catalogued and missing objects,
\begin{equation}
\label{eqS:GALsplit}
\frac{\dd N_{\rm GAL}}{\dd z\,\dd\Omega\,\dd\mathcal G}
=
\frac{\dd N_{\rm cat}}{\dd z\,\dd\Omega\,\dd\mathcal G}
+
\frac{\dd N_{\rm miss}}{\dd z\,\dd\Omega\,\dd\mathcal G}.
\end{equation}
The complete-catalog case is recovered as the limit in which the missing contribution vanishes.

In the present application we take $\mathcal G=L$, the galaxy luminosity derived from observed apparent magnitude $m$, and
\begin{equation}
\label{eqS:weight}
w(L,\epsilon)\propto\left(\frac{L}{L_*}\right)^\epsilon,
\end{equation}
where $L_*$ is the characteristic luminosity of the adopted Schechter luminosity function. Thus $\lambda_g=\epsilon$. The normalization of Eq.~\eqref{eqS:weight} is immaterial because only ratios of galaxy weights enter the inference.

\subsection{Missing galaxies and completeness}
\label{sec:completion}

We adopt homogeneous completion~\cite{Finke:2021aom}: at fixed redshift, missing galaxies are described by the underlying Schechter function and are taken to be homogeneous in comoving volume and sky position, with departures from full catalog coverage encoded by the probability that a galaxy is included in the survey, $P_{\rm det,gal} \in [0,1]$. We write
\begin{equation}
\label{eqS:Nmiss}
\frac{\dd N_{\rm miss}}{\dd z\,\dd\Omega\,\dd\mathcal G}
=
\frac{\dd V_c}{\dd z\,\dd\Omega}(z;\lambda)
\Phi(\mathcal G,z;\lambda_{\rm Sc})
\left[1-P_{\rm det,gal}(\mathcal G,z,\Omega)\right],
\end{equation}
where $\Phi$ is a a generic Schechter-type function of galaxy properties, parameterized here by Schechter parameters $\lambda_{\rm Sc}$, and $V_c$ is the comoving volume. The Schechter parameters are treated as known in the present analysis; their uncertainty is discussed in the main text as an extension of the framework.

For a small region $\mathcal S$ around $(z,\Omega)$, the weighted density of detected galaxies can be estimated directly from the catalog,
\begin{equation}
\label{eqS:detdensity}
\int\dd\mathcal G\,w(\mathcal G;\lambda_g)
\Phi(\mathcal G,z;\lambda_{\rm Sc})P_{\rm det,gal}
\simeq
\frac{1}{V_c(\mathcal S)}
\sum_{g\in\mathcal S}w(\mathcal G_g;\lambda_g).
\end{equation}
This motivates defining the weighted completeness
\begin{equation}
\label{eqS:compgeneral}
P_{\rm comp}(z,\Omega,\lambda_g;\Dgal)
=
\min\!\left[
\frac{\displaystyle V_c(\mathcal S)^{-1}\sum_{g\in\mathcal S}w(\mathcal G_g;\lambda_g)}
{\displaystyle \int\dd\mathcal G\,w(\mathcal G;\lambda_g)\Phi(\mathcal G,z;\lambda_{\rm Sc})},
1\right].
\end{equation}
For the luminosity model of Eq.~\eqref{eqS:weight}, define
\begin{equation}
\label{eqS:lbar}
\bar\ell_{{\rm gal},\epsilon}(z,\epsilon)
\equiv
\int_{L_{\rm min}}^{L_{\rm max}}\dd L\,
w(L,\epsilon)\Phi(L,z;\lambda_{\rm Sc})
\end{equation}
and
\begin{equation}
\label{eqS:lcat}
\ell_{{\rm cat},\epsilon}(z,\Omega,\epsilon;\Dgal)
\equiv
\frac{1}{V_c(\mathcal S)}
\sum_{g\in\mathcal S}w(L_g,\epsilon).
\end{equation}
Note that, in case the Schechter function does not evolve with redshift, $\bar{\ell}_{{\rm gal},\epsilon}$ is constant in redshift. 

The completeness used in the analysis is therefore
\begin{equation}
\label{eqS:pcomp}
P_{\rm comp}(z,\Omega,\epsilon;\Dgal)
=
\min\!\left[
\frac{\ell_{{\rm cat},\epsilon}(z,\Omega,\epsilon;\Dgal)}
{\bar\ell_{{\rm gal},\epsilon}(z, \epsilon)},1\right].
\end{equation}
Because the catalog is weighted with the same $w(L,\epsilon)$ that enters the merger model, the completeness itself depends on the inferred parameter $\epsilon$. This dependence is retained throughout the inference.

In practice, galaxy luminosities are not observed directly but are inferred from apparent magnitudes by assuming a cosmological distance--redshift relation. Furthermore, both $V_c(\mathcal S)$ and the Schechter parameters carry a dependence on cosmology. However, the explicit $H_0$ dependence cancels from the ratio in Eq.~\eqref{eqS:pcomp}: $M_*-M$ is independent of $H_0$, while the Schechter normalization scales as $H_0^3$, matching the comoving-volume scaling. A residual dependence on other cosmological parameters can remain only through the apparent-to-absolute magnitude conversion. Since we fix $\Omega_{\rm m}$ and assume a $\Lambda$CDM cosmology, throughout this work we neglect this residual dependence. Note that the same issue is absent for weights based on galaxy properties that do not require a cosmology-dependent luminosity conversion, such as stellar mass.

Finally, note that the definition of completeness in Eq.~\ref{eqS:compgeneral}, introduced by Ref.~\cite{Finke:2021aom} and used in the dark-siren package {\sc chimera}~\cite{Borghi:2023opd}, is different than the one in the codes {\sc icarogw}~\cite{Mastrogiovanni:2023zbw} and {\sc gwcosmo}~\cite{Gray:2023wgj}, which integrate a Schechter function above an effective, per–pixel magnitude threshold. Instead, we compare the catalog’s weighted density directly to the expected total. This avoids assuming the survey is purely magnitude-limited---or that other selections can be reduced to an effective magnitude cut---and permits completeness to be modeled in full generality including more general selection cuts.

\subsection{Catalog and missing-galaxy distributions}

The observed catalog contains galaxies with measured redshift, sky position, and luminosity $\{\hat z_g,\hat\Omega_g,\hat L_g\}$. We assume negligible sky-position and luminosity uncertainties,  a Gaussian likelihood for the measured redshift with standard deviation $\hat\sigma_{z,g}$, and a unoiform--in--comoving--volume redshift prior. Specifically, define the dimensionless comoving-volume element
\begin{equation}
\label{eqS:jz}
j(z;\lambda)\equiv
\frac{H_0^3}{c^3}\frac{\dd V_c}{\dd z\,\dd\Omega}(z;\lambda),
\end{equation}
that does not depend on $H_0$, and the normalization
\begin{equation}
\label{eqS:Zg}
\mathcal Z_g
\equiv
\int\dd z\,\mathcal N(z;\hat z_g,\hat\sigma_{z,g}^2)\,j(z;\lambda).
\end{equation}
The normalized contribution of catalog galaxies is~\cite{Finke:2021aom,Mastrogiovanni:2023zbw}
\begin{equation}
\label{eqS:pcat}
p_{\rm cat}(z,\Omega\mid\lambda,\Dgal)
= 
\frac{j(z;\lambda)}{\sum_{g\in{\rm cat}}w(\hat L_g,\epsilon)}
\sum_{g\in{\rm cat}}w(\hat L_g,\epsilon)
\,
\frac{\mathcal N(z;\hat z_g,\hat\sigma_{z,g}^2)}{\mathcal Z_g}
\delta(\Omega-\hat\Omega_g).
\end{equation}
For later use, it is convenient to keep the galaxy label explicit and define the corresponding joint catalog density
\begin{equation}
\label{eqS:pcatjoint}
p_{\rm cat}(z,\Omega,g\mid\lambda,\Dgal)
=
\frac{w(\hat L_g,\epsilon)}
{\displaystyle\sum_{g'\in{\rm cat}}w(\hat L_{g'},\epsilon)}
\frac{\mathcal N(z;\hat z_g,\hat\sigma_{z,g}^2)\,j(z;\lambda)}
{\mathcal Z_g}
\delta(\Omega-\hat\Omega_g),
\end{equation}
so that $\sum_{g\in{\rm cat}}p_{\rm cat}(z,\Omega,g\mid\lambda,\Dgal)
=p_{\rm cat}(z,\Omega\mid\lambda,\Dgal)$. It is worth noticing that this construction can be easily generalized to non-Gaussian likelihood hypotheses.

Let $R$ denote the redshift--sky region over which the galaxy distribution is normalized; in practice it is chosen to enclose the GW detector horizon and the analyzed sky region. The weighted catalog fraction is
\begin{equation}
\label{eqS:fR}
f_R(\epsilon;\Dgal)
=
\frac{\displaystyle\int_R\dd z\,\dd\Omega\,
j(z;\lambda)\bar\ell_{{\rm gal},\epsilon}(z,\epsilon)
P_{\rm comp}(z,\Omega,\epsilon;\Dgal)}
{\displaystyle\int_R\dd z\,\dd\Omega\,
j(z;\lambda)\bar\ell_{{\rm gal},\epsilon}(z,\epsilon)}.
\end{equation}
The corresponding normalized missing-galaxy distribution is
\begin{equation}
\label{eqS:pmiss}
p_{\rm miss}(z,\Omega\mid\lambda,\Dgal)
={}
\frac{1-P_{\rm comp}(z,\Omega,\epsilon;\Dgal)}{1-f_R(\epsilon;\Dgal)} \,
\frac{j(z;\lambda)\bar\ell_{{\rm gal},\epsilon}(z,\epsilon)}
{\displaystyle\int_R\dd z\,\dd\Omega\,
j(z;\lambda)\bar\ell_{{\rm gal},\epsilon}(z,\epsilon)}.
\end{equation}
Catalogued and missing galaxies are finally combined as
\begin{equation}
\label{eqS:pgal}
p_{\rm gal}(z,\Omega\mid\lambda,\Dgal)
={}
f_R(\epsilon;\Dgal)\,p_{\rm cat}(z,\Omega\mid\lambda,\Dgal)
+\left[1-f_R(\epsilon;\Dgal)\right]\,
p_{\rm miss}(z,\Omega\mid\lambda,\Dgal).
\end{equation}
The source redshift--sky distribution entering Eq.~\eqref{eqS:popfactor} is therefore proportional to $\psi(z\mid\lambda)/(1+z)$ times Eq.~\eqref{eqS:pgal}. For a complete catalog, $P_{\rm comp}=f_R=1$, the missing component vanishes, and $p_{\rm gal}=p_{\rm cat}$.

\section{Full-posterior sampling and numerical implementation}
\label{secS:sampling}

We now provide the details of the full-posterior sampling strategy
summarized in the main text. The central idea is to retain the
event-level parameters, including the discrete host variables, rather
than marginalizing them within the hierarchical likelihood. This makes
the posterior probabilities of individual host galaxies and of the
out-of-catalog hypothesis direct outputs of the inference.

\subsection{General sampling formulation}
\label{secS:general_sampling}

For $N_{\rm ev}$ detections, the joint posterior for the event-level
parameters $\Theta\equiv\{\theta_i\}_{i=1}^{N_{\rm ev}}$ and the
hyperparameters $\lambda$ can be written as
\begin{equation}
\label{eqS:full_posterior}
p(\Theta,\lambda\mid\DGW,\Dgal)
\propto
\frac{\pi(\lambda)}{\xi(\lambda)^{N_{\rm ev}}}
\prod_{i=1}^{N_{\rm ev}}
\mathcal L(d_i\mid\theta_i)\,
p_{\rm pop}(\theta_i\mid\lambda,\Dgal),
\end{equation}
where $\mathcal L(d_i\mid\theta_i)$ is the single-event GW likelihood,
$\pi(\lambda)$ is the hyperprior, and
\begin{equation}
\label{eqS:selection}
\xi(\lambda)
=
\int \dd\theta\,
P_{\rm det}(\theta)\,
p_{\rm pop}(\theta\mid\lambda,\Dgal)
\end{equation}
is the fraction of detectable events in the population. As in the main
text, the overall merger rate has been marginalized using a
scale-invariant prior.

Following Ref.~\cite{Mancarella:2025uat}, Eq.~\eqref{eqS:full_posterior}
can be written more generally by introducing an arbitrary normalized
sampling proposal
$\pi(\theta_i\mid d_i,\Dgal,\lambda)$ for the event-level parameters:
\begin{equation}
\label{eqS:full_posterior_with_proposal}
p(\Theta,\lambda\mid\DGW,\Dgal)
\propto{}
\pi(\lambda)
\left[
\prod_{i=1}^{N_{\rm ev}}
\pi(\theta_i\mid d_i,\Dgal,\lambda)
\right]
\,
\xi(\lambda)^{-N_{\rm ev}}
\prod_{i=1}^{N_{\rm ev}}
\mathcal L(d_i\mid\theta_i)\,
\frac{
p_{\rm pop}(\theta_i\mid\lambda,\Dgal)
}{
\pi(\theta_i\mid d_i,\Dgal,\lambda)
}.
\end{equation}
The proposal $\pi(\theta_i\mid d_i,\Dgal,\lambda)$ is a computational
sampling distribution and should not be confused with the
parameter-estimation prior $\pi_{\rm PE}(\theta_i)$. Equation
\eqref{eqS:full_posterior_with_proposal} is simply Eq.~\eqref{eqS:full_posterior}
multiplied and divided by the same normalized proposal. It therefore
allows us to choose a distribution that efficiently samples the
event-level posterior, while the ratio between the population target and
the proposal provides the corresponding correction.

In the absence of galaxy information, a natural choice is the
single-event GW posterior,
$\pi(\theta_i\mid d_i,\lambda)=p(\theta_i\mid d_i,\lambda)$, so that
the product of single-event posteriors acts as the sampling distribution
over the event-level parameters~\cite{Mancarella:2025uat}. In the
present case we instead construct a proposal that combines the GW
posterior with the galaxy information, as described below.

\subsection{Gaussian-mixture representation of the GW posteriors}
\label{secS:gmm}

For the purpose of sampling, we distinguish the detector-frame
parameters directly constrained by the GW data from the corresponding
source-frame quantities entering the population model. We represent
each single-event posterior in the transformed detector-frame variables
\begin{equation}
\label{eqS:theta_tilde}
\tilde{\theta}_{\rm D}
=
\left\{
\log\mathcal{M}_{z},\,
\log\frac{q}{1-q},\,
\log d_L,\,
\log\frac{\alpha}{2\pi-\alpha},\,
\log\frac{\delta+\pi/2}{\pi/2-\delta}
\right\},
\end{equation}
where
\begin{equation}
\mathcal{M}_{z}
=
\frac{
(m_{1,{\rm D}}m_{2,{\rm D}})^{3/5}
}{
(m_{1,{\rm D}}+m_{2,{\rm D}})^{1/5}
}
\end{equation}
is the detector-frame chirp mass,
$q=m_{2,{\rm D}}/m_{1,{\rm D}}$ is the mass ratio,
$d_L$ is the luminosity distance, and
$\{\alpha,\delta\}\equiv\Omega$ are right ascension and declination.
The transformations in Eq.~\eqref{eqS:theta_tilde} map the variables
onto unbounded domains and avoid sharp boundaries in the posterior
interpolation.

We approximate each single-event posterior with a Gaussian Mixture
Model (GMM), following Ref.~\cite{Mancarella:2025uat}. The number of
components is selected by minimizing the Bayesian Information Criterion
(BIC); after locating its minimum, we verify that adding at least ten
additional components does not improve the BIC.

For event $i$, we introduce the auxiliary GMM-component index
\begin{equation}
\label{eqS:gmm_indices}
k_i
\sim
\pi(k_i\mid d_i)
=
\mathcal C(w_{k,i})
\in
\{1,\ldots,N_{\rm gmm}^i\},
\end{equation}
where $\mathcal C$ denotes a categorical distribution,
$w_{k,i}$ are the normalized GMM weights, and $N_{\rm gmm}^i$ is the
number of mixture components. The GMM representation factorizes as
\begin{equation}
\label{eqS:gmm_factorization}
p(\theta_i,k_i\mid d_i,\lambda)
=
p(\theta_i\mid k_i,d_i,\lambda)\,
\pi(k_i\mid d_i),
\end{equation}
and marginalizing over $k_i$ recovers the GMM approximation to the
single-event posterior.

In practice, we sample the detector-frame source parameters
$\bar\theta_{{\rm D},i}$ from the GMM posterior marginalized over
luminosity distance and sky position. Once a redshift $z_i$ and sky
position $\Omega_i$ have been selected from the galaxy-informed
proposal described below, the luminosity distance is fixed by the
cosmological parameters,
\begin{equation}
\label{eqS:dl_from_z}
d_{L,i}=d_L(z_i;\lambda_{\rm c}),
\end{equation}
and the full GMM posterior is evaluated at
$(\bar\theta_{{\rm D},i},d_{L,i},\Omega_i)$. The detector-frame masses
are then converted to source-frame masses using $z_i$ before evaluating
the population model.

Explicitly, the proposal density for the corresponding source-frame
parameters $\bar\theta_i$ is
\begin{equation}
\label{eqS:source_proposal}
p(\bar\theta_i\mid k_i,d_i,z_i)
={}
p(\bar\theta_{{\rm D},i}\mid k_i,d_i)
\left|
\frac{\dd\bar\theta_{{\rm D},i}}
     {\dd\bar\theta_i}
\right|
={}
(1+z_i)^2\,
p(\bar\theta_{{\rm D},i}\mid k_i,d_i),
\end{equation}
where the last equality follows from the transformation of the two
component masses.

Up to the single-event evidence, which is independent of the sampled
parameters, the GW likelihood is related to the PE posterior through
\begin{equation}
\label{eqS:single_event_likelihood}
\mathcal L(d_i\mid\theta_i,k_i,\lambda_{\rm c})\propto
\frac{
p(\theta_i,k_i\mid d_i,\lambda_{\rm c})
}{
\pi_{\rm PE}(\theta_i\mid\lambda_{\rm c})
}=
\frac{
p(\theta_i\mid k_i,d_i,\lambda_{\rm c})\,
\pi(k_i\mid d_i)
}{
\pi_{\rm PE}(\theta_i\mid\lambda_{\rm c})
}.
\end{equation}
Importantly, the GW likelihood itself contains no galaxy-catalog
information; the latter enters through the population model and the
sampling proposal.

For the PE priors used here, which are flat in detector-frame component
masses, the corresponding prior in source-frame variables is
\begin{equation}
\label{eqS:PE_prior}
\pi_{\rm PE}(\theta_i\mid\lambda_{\rm c})
=
(1+z_i)^2
\frac{\partial d_L}{\partial z}(z_i;\lambda_{\rm c})\,
\pi_{\rm PE}(d_{L,i}),
\end{equation}
where $\pi_{\rm PE}(d_L)$ is the luminosity-distance prior used in the
single-event parameter estimation.

\subsection{Catalog and missing-galaxy branches}
\label{secS:branches}

The full-posterior formulation is particularly useful here because the
variables describing the host association are themselves inference
targets. We therefore retain explicitly, for every event, a binary
variable $s_i$ indicating whether the host is represented in the
catalog and, on the catalog branch, a discrete host-galaxy label $g_i$.
The resulting hierarchical posterior is the one introduced in the main
text,
\begin{equation}
\label{eqS:full_posterior_discrete}
p(\Theta,\bm g,\bm s,\lambda\mid\DGW,\Dgal)\propto \frac{\pi(\lambda)}{\xi(\lambda)^{N_{\rm ev}}}\prod_{i=1}^{N_{\rm ev}}\mathcal L(d_i\mid\theta_i)\,p_{\rm pop}(\theta_i,g_i,s_i\mid\lambda,\Dgal).
\end{equation}
Here $\bm g\equiv\{g_i\}_{i=1}^{N_{\rm ev}}$ and
$\bm s\equiv\{s_i\}_{i=1}^{N_{\rm ev}}$. Marginalization over the
auxiliary GMM-component indices introduced previously is implicit throughout this section, since those
indices are only used internally by the sampler and are not inference
products of interest.

The variable $s_i$ explicitly unpacks the catalog and missing-galaxy
mixture derived in Sec.~\ref{secS:hierarchy}. We assign
\begin{equation}
\label{eqS:switch}
p(s_i=1\mid\lambda,\Dgal)=f_R(\epsilon;\Dgal),\qquad p(s_i=0\mid\lambda,\Dgal)=1-f_R(\epsilon;\Dgal),
\end{equation}
where $s_i=1$ denotes the hypothesis that the host is represented in
the catalog and $s_i=0$ the hypothesis that it belongs to the
missing-galaxy population. When $s_i=1$, $g_i$ labels one of the
candidate host galaxies. When $s_i=0$, the value of $g_i$ is an
auxiliary variable with no physical interpretation and can be assigned
any normalized distribution, which cancels from the posterior.

Conditioned on the branch, the galaxy-dependent part of the population
distribution reduces to the corresponding component of the mixture in
Sec.~\ref{secS:hierarchy}. The catalog branch contains
$p_{\rm cat}(z_i,\Omega_i,g_i\mid\lambda,\Dgal)$, where the explicit
label $g_i$ selects one term of the catalog mixture, while the
out-of-catalog branch contains
$p_{\rm miss}(z_i,\Omega_i\mid\lambda,\Dgal)$.

\subsection{Sampling proposal and proposal-to-target correction}
\label{secS:proposal}

Equation~\eqref{eqS:full_posterior_discrete} defines the target
distribution, but it need not be sampled directly. As in
Eq.~\eqref{eqS:full_posterior_with_proposal}, we introduce an arbitrary
normalized proposal for the event-level variables and write
\begin{equation}
\label{eqS:full_posterior_discrete_proposal}
p(\Theta,\bm g,\bm s,\lambda\mid\DGW,\Dgal)\propto \pi(\lambda)\left[\prod_{i=1}^{N_{\rm ev}}\pi(\theta_i,g_i,s_i\mid d_i,\Dgal,\lambda)\right]\xi(\lambda)^{-N_{\rm ev}}\prod_{i=1}^{N_{\rm ev}}\frac{\mathcal L(d_i\mid\theta_i)\,p_{\rm pop}(\theta_i,g_i,s_i\mid\lambda,\Dgal)}{\pi(\theta_i,g_i,s_i\mid d_i,\Dgal,\lambda)}.
\end{equation}
The proposal can therefore be chosen for computational convenience,
provided that the corresponding proposal-to-target ratio in
Eq.~\eqref{eqS:full_posterior_discrete_proposal} is retained exactly.

Our choice is designed as a compromise between being straightforward
to sample and remaining close to the two factors defining the target.
The source parameters directly constrained by the GW data are proposed
from the single-event GW posterior marginalized over luminosity
distance and sky position, while redshift and sky position are proposed
from the appropriate galaxy branch. Explicitly,
\begin{equation}
\label{eqS:proposal_factorization}
\pi(\theta_i,g_i,s_i\mid d_i,\Dgal,\lambda)=p(\bar\theta_i\mid d_i,z_i)\,\pi(z_i,\Omega_i,g_i\mid s_i,\Dgal,\lambda)\,\pi(s_i\mid\Dgal,\lambda),
\end{equation}
where $p(\bar\theta_i\mid d_i,z_i)$ is the source-frame proposal induced
by the GMM approximation to the detector-frame posterior marginalized
over luminosity distance and sky position. The GMM-component
marginalization is implicit.

We choose the branch proposal to coincide with its population
probability,
\begin{equation}
\label{eqS:branch_proposal}
\pi(s_i=1\mid\Dgal,\lambda)=f_R(\epsilon;\Dgal),\qquad \pi(s_i=0\mid\Dgal,\lambda)=1-f_R(\epsilon;\Dgal).
\end{equation}
The branch probability therefore cancels exactly between target and
proposal. The remaining correction consists of a GW-posterior factor,
common to both branches, the compact-binary population factor, and a
galaxy factor that depends on whether $s_i=0$ or $s_i=1$. Schematically,
for either branch,
\begin{equation}
\label{eqS:master_correction}
\frac{\mathcal L(d_i\mid\theta_i)\,p_{\rm pop}(\theta_i,g_i,s_i\mid\lambda,\Dgal)}{\pi(\theta_i,g_i,s_i\mid d_i,\Dgal,\lambda)}\propto \frac{\mathcal L(d_i\mid\theta_i)}{p(\bar\theta_i\mid d_i,z_i)}\,p_{\rm pop}(\bar\theta_i\mid\lambda)\,\frac{\psi(z_i\mid\lambda_{\rm a})}{1+z_i}\,\frac{p_{\rm gal}(z_i,\Omega_i,g_i\mid s_i,\lambda,\Dgal)}{\pi(z_i,\Omega_i,g_i\mid s_i,\Dgal,\lambda)},
\end{equation}
where in the last ratio
$p_{\rm gal}=p_{\rm miss}$ for $s_i=0$ and
$p_{\rm gal}=p_{\rm cat}$ for $s_i=1$. For $s_i=0$, the dependence on
the auxiliary $g_i$ is understood to cancel and is omitted below.

Equation~\eqref{eqS:master_correction} makes explicit the logic of the
sampling construction. Whenever the proposal is chosen to coincide
with the corresponding part of the population target, no correction is
required. Where this is not computationally convenient, the residual
target-to-proposal ratio is evaluated explicitly.

\subsection{GW-posterior contribution to the correction}
\label{secS:GWfactor}

The first factor on the right-hand side of
Eq.~\eqref{eqS:master_correction} is common to the catalog and
out-of-catalog branches. Up to the single-event evidence,
\begin{equation}
\label{eqS:GW_common}
\frac{\mathcal L(d_i\mid\theta_i)}{p(\bar\theta_i\mid d_i,z_i)}\propto \frac{1}{\pi_{\rm PE}(\theta_i\mid\lambda_{\rm c})}\frac{p(\theta_i\mid d_i,\lambda_{\rm c})}{p(\bar\theta_i\mid d_i,z_i)}.
\end{equation}
The numerator is the full GMM approximation evaluated at the
luminosity distance and sky position selected by the active galaxy
branch, whereas the denominator is the marginalized GW proposal from
which $\bar\theta_i$ is sampled.

Using the transformed variables introduced in
Sec.~\ref{secS:gmm}, their ratio is
\begin{equation}
\label{eqS:GW_ratio}
\frac{p(\theta_i\mid d_i,\lambda_{\rm c})}{p(\bar\theta_i\mid d_i,z_i)}=R_{p,i}\,\frac{1}{d_{L,i}}\,\frac{\partial d_L}{\partial z}(z_i;\lambda_{\rm c})\,J_{\Omega,i},
\end{equation}
where
\begin{equation}
\label{eqS:Rp}
R_{p,i}\equiv\frac{p(\tilde\theta_{{\rm D},i}\mid d_i,\lambda_{\rm c})}{p(\tilde{\bar\theta}_{{\rm D},i}\mid d_i)},
\qquad
J_{\Omega,i}\equiv\left|\frac{2\pi}{\alpha_i(2\pi-\alpha_i)}\right|\left|\frac{\pi}{\pi^2/4-\delta_i^2}\right|.
\end{equation}
Here $\tilde{\bar\theta}_{{\rm D},i}$ denotes the subset of transformed
detector-frame variables retained after marginalizing over luminosity
distance and sky position.

For the parameter-estimation priors adopted in this work,
\begin{equation}
\label{eqS:PE_prior_branch}
\pi_{\rm PE}(\theta_i\mid\lambda_{\rm c})=(1+z_i)^2\,\frac{\partial d_L}{\partial z}(z_i;\lambda_{\rm c})\,\pi_{\rm PE}(d_{L,i}),
\end{equation}
where $\pi_{\rm PE}(d_L)$ is the luminosity-distance prior used for the
single-event analysis. Equations~\eqref{eqS:GW_common}--\eqref{eqS:PE_prior_branch}
fully specify the GW contribution that multiplies the branch-dependent
population-to-proposal factors discussed next.

\subsection{Out-of-catalog branch}
\label{secS:missing_branch}

For $s_i=0$, the natural choice is to sample directly from the
missing-galaxy distribution,
\begin{equation}
\label{eqS:missing_proposal}
\pi(z_i,\Omega_i\mid s_i=0,\Dgal,\lambda)=p_{\rm miss}(z_i,\Omega_i\mid\Dgal,\lambda).
\end{equation}
The galaxy part of the final ratio in
Eq.~\eqref{eqS:master_correction} therefore cancels identically,
\begin{equation}
\label{eqS:missing_ratio}
\frac{p_{\rm miss}(z_i,\Omega_i\mid\Dgal,\lambda)}{\pi(z_i,\Omega_i\mid s_i=0,\Dgal,\lambda)}=1.
\end{equation}
The complete proposal-to-target correction for this branch is thus
proportional to
\begin{equation}
\label{eqS:missing_final}
\frac{\mathcal L(d_i\mid\theta_i)\,p_{\rm pop}(\theta_i,g_i,s_i=0\mid\lambda,\Dgal)}{\pi(\theta_i,g_i,s_i=0\mid d_i,\Dgal,\lambda)}\propto \frac{\mathcal L(d_i\mid\theta_i)}{p(\bar\theta_i\mid d_i,z_i)}\,p_{\rm pop}(\bar\theta_i\mid\lambda)\,\frac{\psi(z_i\mid\lambda_{\rm a})}{1+z_i}.
\end{equation}
Thus the missing-galaxy branch is particularly simple: its redshift and
sky position are proposed from exactly the same distribution that
appears in the population model, leaving only the common GW and
compact-binary population factors.

\subsection{In-catalog branch and event-specific proposal}
\label{secS:catalog_branch}

For $s_i=1$, one could in principle sample directly from the full
catalog distribution $p_{\rm cat}$. In practice this is inefficient,
because the overwhelming majority of catalog galaxies lie outside the
region supported by the GW likelihood. We therefore restrict the
sampling proposal to an event-specific candidate set
${\rm cat}_i^{\rm in}\subset{\rm cat}$.

The set ${\rm cat}_i^{\rm in}$ contains galaxies whose measured sky
positions lie inside an adopted localization region $\Delta\Omega_i$  and whose measured redshifts lie inside an interval $\Delta z_i$ compatible with the GW luminosity-distance posterior. The
redshift interval is constructed from the endpoints of the $H_0$ prior and a $\pm n\sigma$ expansion of the GW luminosity-distance uncertainty. In our simulation study, we require the measured sky position to lie within the $90\%$ credible region of the GW sky localization, with a redshift window corresponding to $n=3$.


Within this set, we use the catalog proposal
\begin{equation}
\label{eqS:catalog_proposal}
\pi(z_i,\Omega_i,g_i\mid s_i=1,\Dgal,\lambda)\equiv\pi_{\rm in}(z_i,\Omega_i,g_i)=\frac{w_{g_i}}{\sum_{g\in{\rm cat}_i^{\rm in}}w_g}\,\mathcal N(z_i;\hat z_{g_i},\hat\sigma_{z,g_i}^2)\,\delta(\Omega_i-\hat\Omega_{g_i}),
\end{equation}
with
$w_{g_i}=w(\hat L_{g_i},\epsilon)$. This proposal is deliberately close
to the catalog population distribution while remaining computationally
tractable. It does not, however, coincide exactly with
$p_{\rm cat}$, because it is normalized only over the event-specific
candidate set and because the catalog redshift distribution contains
the comoving-volume prior derived in Sec.~\ref{secS:hierarchy}.

Consequently, the catalog branch retains the residual correction
\begin{equation}
\label{eqS:catalog_master}
\frac{\mathcal L(d_i\mid\theta_i)\,p_{\rm pop}(\theta_i,g_i,s_i=1\mid\lambda,\Dgal)}{\pi(\theta_i,g_i,s_i=1\mid d_i,\Dgal,\lambda)}\propto \frac{\mathcal L(d_i\mid\theta_i)}{p(\bar\theta_i\mid d_i,z_i)}\,p_{\rm pop}(\bar\theta_i\mid\lambda)\,\frac{\psi(z_i\mid\lambda_{\rm a})}{1+z_i}\,\frac{p_{\rm cat}(z_i,\Omega_i,g_i\mid\lambda,\Dgal)}{\pi_{\rm in}(z_i,\Omega_i,g_i)}.
\end{equation}

For a specific galaxy $g_i$, the redshift factor entering the catalog
population distribution is
\begin{equation}
\label{eqS:single_galaxy_target}
p_{\rm cat}(z_i\mid g_i, \lambda)=\frac{\mathcal N(z_i;\hat z_{g_i},\hat\sigma_{z,g_i}^2)\,j(z_i; \lambda)}{\mathcal Z_{g_i}},
\end{equation}
Ignoring for the moment the common sky-position delta function, the
remaining catalog correction is therefore
\begin{equation}
\label{eqS:catalog_correction}
\frac{p_{\rm cat}(z_i,\Omega_i,g_i\mid\lambda,\Dgal)}{\pi_{\rm in}(z_i,\Omega_i,g_i)}\simeq \frac{j(z_i;\lambda)}{\mathcal Z_{g_i}}\,\frac{\sum_{g\in{\rm cat}_i^{\rm in}}w_g}{\sum_{g\in{\rm cat}}w_g}.
\end{equation}
The first factor accounts for the comoving-volume prior entering the
redshift distribution of each catalog galaxy; the second corrects the
different normalization of the event-specific proposal and the full
catalog population.

The normalization correction is particularly important when
$\epsilon$ is inferred. If the galaxy weights are fixed and the
candidate set contains essentially all the relevant GW support,
approximating the restricted proposal as the catalog population can
have little effect. When $\epsilon$ is sampled, however, both sums in
Eq.~\eqref{eqS:catalog_correction} depend on $\epsilon$, and the
correction accumulates over all events.
We verified that neglecting it can bias the recovered luminosity-weighting parameter, while including Eq.~\eqref{eqS:catalog_correction} removes this effect.

The contrast between the two branches illustrates the purpose of the
proposal construction. For missing galaxies we can choose a proposal
that coincides exactly with the population distribution, so the galaxy
correction vanishes. For catalog galaxies, a restricted proposal is
required for computational efficiency, and the corresponding
$p_{\rm cat}/\pi_{\rm in}$ factor restores the exact hierarchical
target.

\subsection{Sampling algorithm}
\label{secS:algorithm}

The resulting sampling procedure can be summarized as follows:
\begin{enumerate}[label=(\roman*)]

\item Sample the population and cosmological hyperparameters
$\lambda$, including $\epsilon$.

\item For each event $i=1,\ldots,N_{\rm ev}$:
\begin{enumerate}[label=(\alph*)]

\item sample the detector-frame source parameters
$\bar\theta_{{\rm D},i}$ from the GMM representation of the GW
posterior marginalized over luminosity distance and sky position;

\item sample an in-catalog candidate
$(g_i,z_i^{\rm cat},\Omega_i^{\rm cat})$ from
$\pi_{\rm in}$ in Eq.~\eqref{eqS:catalog_proposal};

\item sample
$(z_i^{\rm miss},\Omega_i^{\rm miss})$ from
$p_{\rm miss}$;

\item sample
$s_i\sim{\rm Bernoulli}[f_R(\epsilon;\Dgal)]$;

\item select
\begin{equation}
\label{eqS:active_branch}
(z_i,\Omega_i)=\begin{cases}(z_i^{\rm cat},\Omega_i^{\rm cat}),&s_i=1,\\(z_i^{\rm miss},\Omega_i^{\rm miss}),&s_i=0,\end{cases}
\end{equation}
compute $d_{L,i}=d_L(z_i;\lambda_{\rm c})$, evaluate the full GW
posterior at $(\bar\theta_{{\rm D},i},d_{L,i},\Omega_i)$, and transform
the detector-frame source parameters to the corresponding source-frame
parameters using $z_i$.

\end{enumerate}

\item Evaluate the hierarchical posterior
Eq.~\eqref{eqS:full_posterior_discrete}, including the common GW
correction and the branch-dependent proposal-to-target factors derived
above.

\end{enumerate}

Both branches are sampled for every event, but only the branch selected
by $s_i$ enters the GW likelihood and population model. The inactive
branch is an auxiliary draw from a normalized distribution and
integrates out of the posterior.

\subsection{Host probabilities}
\label{secS:host_probabilities}

Because $g_i$ and $s_i$ are retained explicitly in the full posterior,
the host probabilities are obtained directly from the posterior samples
and are automatically marginalized over the compact-binary population,
the galaxy--merger relation, and cosmology.

For an incomplete catalog, the probability of galaxy $g$ conditional
on the host being represented in the catalog is
\begin{equation}
\label{eqS:pcond}
p_{g,i}^{\rm cond}\equiv p(g_i=g\mid s_i=1,\DGW,\Dgal).
\end{equation}
The absolute probability that galaxy $g$ is the true host is instead
\begin{equation}
\label{eqS:pabs}
p_{g,i}^{\rm abs}\equiv p(s_i=1,g_i=g\mid\DGW,\Dgal)=p_{g,i}^{\rm cond}\,p(s_i=1\mid\DGW,\Dgal).
\end{equation}
The complementary probability, not to be confused with the out-of-catalog component of the redshift prior,
\begin{equation}
\label{eqS:pmiss_event}
p_{{\rm miss},i}\equiv p(s_i=0\mid\DGW,\Dgal)
\end{equation}
is the posterior probability that the true host is absent from the
catalog, so that
\begin{equation}
\label{eqS:host_normalization}
\sum_{g=1}^{N_{\rm gal}^i}p_{g,i}^{\rm abs}+p_{{\rm miss},i}=1.
\end{equation}

In the complete-catalog limit,
$p_{{\rm miss},i}=0$ and
$p_{g,i}^{\rm abs}=p_{g,i}^{\rm cond}=p(g_i=g\mid\DGW,\Dgal)$.
Samples of $g_i$ obtained when $s_i=0$ are purely auxiliary and are
never included in the physical host probabilities.

\subsection{Selection effects and numerical implementation}
\label{secS:selection_implementation}

The selection factor $\xi(\lambda)$ in
Eq.~\eqref{eqS:full_posterior_discrete} is evaluated through reweighted
Monte Carlo integration, as customary in GW-population inference~\cite{Tiwari:2017ndi,Farr:2019rap}. The
same detection and localization criteria used to define the analyzed
GW sample and galaxy catalog are applied to the injection set. For this calculation, the galaxy-informed
distribution derived in Sec.~\ref{secS:hierarchy} is interpolated over
a two-dimensional grid in $(z,\epsilon)$, treating the catalog as a
single effective line of sight.

Drawing $N_{\rm inj}$ injections from $\pi_{\rm inj}(\theta)$ and letting
$\mathcal{D}_{\rm inj}$ denote the subset that survives the same
selection threshold applied to the event catalog, we obtain the
estimator
\begin{equation}
\label{eqS:alpha_inj}
\widehat{\xi}(\lambda)
=
\frac{1}{N_{\rm inj}}
\sum_{k\in\mathcal{D}_{\rm inj}}
\frac{
p_{\rm pop}\!\left(
\theta_k
\mid\lambda,\Dgal
\right)
}{
\pi_{\rm inj}(\theta_k)
}.
\end{equation}
The likelihood variance for the joint posterior in Eq.~\eqref{eqS:full_posterior} is solely related to the finite injection set:
\begin{equation}
\label{eqS:loglik_variance_full}
\sigma^2_{\ln\widehat{\mathcal L}}(\lambda)
\simeq
N_{\rm ev}^2
\frac{
\sigma^2_{\widehat{\xi}}(\lambda)
}{
\widehat{\xi}(\lambda)^2
},
\end{equation}
a quantity we constrain to remain below unity throughout all the analyses.

The inference is implemented in \textsc{PyMC}
~\cite{rainforth2017automating,pymc2023}. Continuous parameters are sampled
with the No-U-Turn Sampler~\cite{JMLR:v15:hoffman14a}, while the discrete
variables $g_i$ and $s_i$ are updated with dedicated categorical and
Bernoulli Gibbs-Metroplis~\cite{article} steps. The GMM-component indices used internally to evaluate
the single-event surrogates are marginalized over 
Convergence, discrete-variable mixing,
selection-function stability, and computational performance are discussed in Sec.~\ref{secS:numerics}.

\section{Simulated galaxy and GW populations}
\label{secS:sims}

\subsection{MICE galaxy catalog}
\label{sec:mice}

We construct the mock galaxy sample from the MICE Grand Challenge light-cone simulation (v2)~\cite{Carretero:2014ltj,Fosalba:2013wxa,Crocce:2013vda,Fosalba:2013mra}. MICE covers one octant of the sky, approximately $5157\,\mathrm{deg}^2$, and was designed to reproduce a DES-like galaxy sample to observed magnitude $i<24$ and redshift $z<1.4$. The simulation assumes a flat $\Lambda$CDM cosmology with $H_0=70\,\kmsMpc$ and $\Omega_m=0.25$.

We retain galaxies in the $^{0.1}\rm r$-band absolute-magnitude interval
\begin{equation}
-21.5\leq M-5\log h_{0.7}\leq-19.5,
\end{equation}
where $h_{0.7}\equiv H_0/(70\,\kmsMpc)$, and in the redshift range $0.07\leq z\leq1.2$. We then subsample the result to obtain a distribution uniform in comoving volume. The parent sample contains approximately $8\times10^7$ galaxies, about $16\%$ of the full MICE catalog.

To evaluate Eq.~\eqref{eqS:pcomp}, we fit a non-evolving Schechter luminosity function to this parent sample. The fitted parameters are
\begin{equation}
\label{eqS:schechterpars}
M_*-5\log h_{0.7}=-20.28,
\qquad
\phi_*h_{0.7}^3=5.78\times10^{-4}\,\mathrm{Mpc}^{-3},
\qquad
\alpha=-0.91.
\end{equation}
They are held fixed in the present analysis. The parent sample is $>95\%$ complete throughout the adopted redshift interval for $\epsilon\in[-3,3]$. We assume spectroscopic redshift uncertainty
\begin{equation}
\sigma_z=0.001(1+z)
\end{equation}
and neglect sky-position and magnitude uncertainties.

To test incomplete catalogs, we apply the apparent-magnitude threshold
\begin{equation}
\label{eqS:mth}
m<m_{\rm th}=21,
\end{equation}
leaving approximately $3.5\times10^6$ galaxies. The threshold is not intended to reproduce a particular survey. It is chosen as a controlled intermediate-completeness case: the catalog remains informative around the peak of the simulated GW redshift distribution while the missing branch becomes important in the high-redshift tail. For $\epsilon=0$, the weighted completeness falls below $50\%$ at $z\simeq0.3$ and is approximately $20\%$ by $z\simeq0.35$. Because the magnitude cut removes faint galaxies preferentially, the completeness has a nontrivial dependence on $\epsilon$ and therefore tests the coupling between the inferred galaxy weighting and the missing-galaxy correction.

\begin{figure}[t]
\centering
\includegraphics[width=0.45\textwidth]{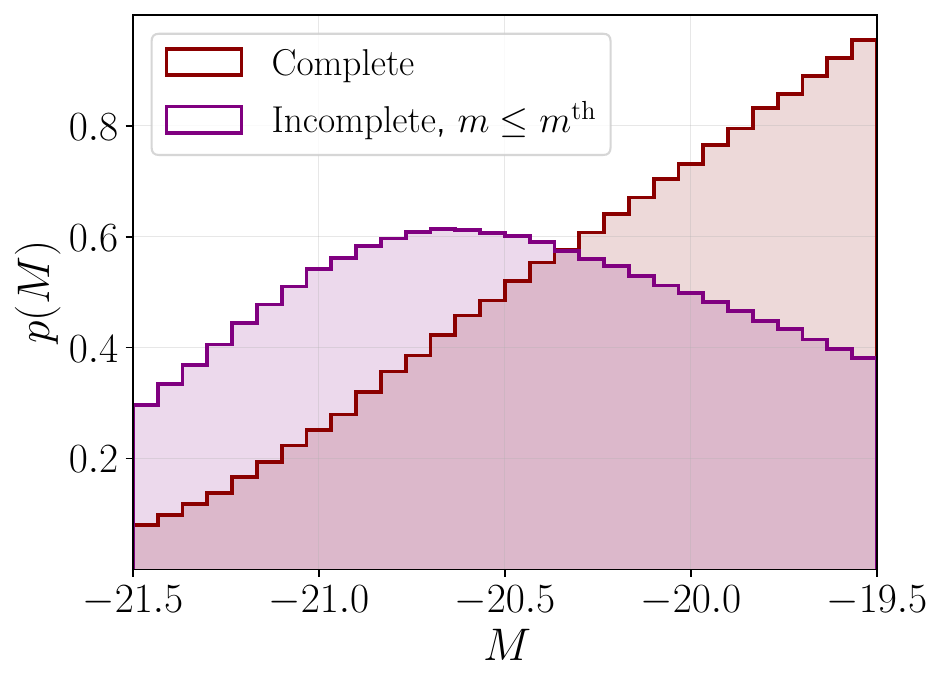}
\hfill
\includegraphics[width=0.45\textwidth]{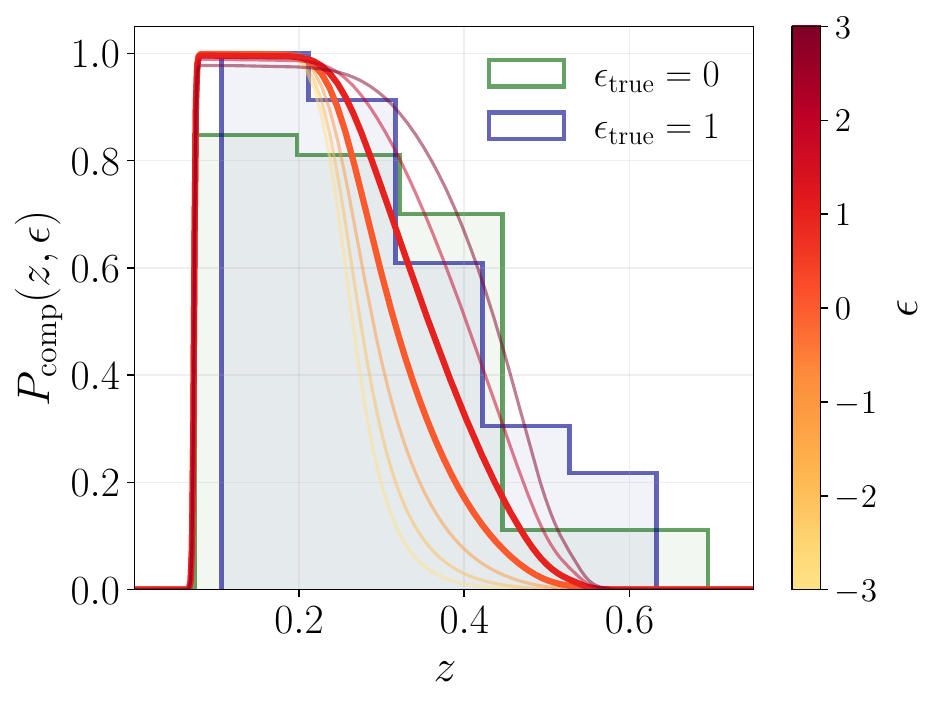}
\caption{Left: $^{0.1}\rm r$-band absolute magnitude distribution of complete and incomplete galaxy catalogs. Right: Catalog completeness as function of redshift and luminosity weight $\epsilon$ over the true redshift distributions of our two different simulated GW catalogs. Thick lines represent completeness at $\epsilon=0$ and $\epsilon=1$.}
\label{figS:micecomp}
\end{figure}

Figure~\ref{figS:micecomp} shows the magnitude distribution (left) of the complete and incomplete catalogs and the catalog completeness of the magnitude--limited catalog (right), superimposed to the redshift distribution of simulated GW events (described in the next section).

We find that the true host galaxy is missing from the galaxy catalog
for a subset of events in our simulated catalogs. Specifically, the true host is absent from the host-candidate galaxies within the
localization volume for 23 events for $\epsilon_{\rm true}=0$ and 20
events for $\epsilon_{\rm true}=1$. The redshift distribution of these
events' posteriors is skewed towards the high-redshift tail of the simulated population, with a median redshift of about $0.4$ and $0.42$ for $\epsilon_{\rm true}=0$ and $1$, consistent with the completeness drop of the magnitude-limited catalog described above. We note that every event retains at least some host-candidate galaxies within its localization
volume, even when the true host itself is not among them.

\subsection{Injected BBH population and GW likelihoods}
\label{sec:gwsim}

We generate two BBH populations with injected host-weight exponents $\epsilon_{\rm true}=0$ and $\epsilon_{\rm true}=1$. Host selection includes a Madau--Dickinson merger-rate evolution~\cite{Madau:2014bja}
\begin{equation}
\label{eqS:MD}
\psi(z\mid\gamma,\kappa,z_p)
=
\frac{(1+z)^\gamma}
{1+\left[(1+z)/(1+z_p)\right]^{\gamma+\kappa}},
\end{equation}
with fiducial values
\begin{equation}
\gamma=3.2,\qquad \kappa=3,\qquad z_p=2.
\end{equation}
For each selected host, component masses are drawn from the ``Broken Power Law + 2 Peaks'' BBH population model used in the current LVK population analysis~\cite{LIGOScientific:2025pvj}. 
The fiducial parameters used to generate the simulated population are listed in Table~\ref{tab:mass_fiducial}, together with the priors adopted in
the inference. The minimum primary and secondary masses,
$m_{1,\min}$ and $m_{2,\min}$, and their taper widths are sampled
ensuring $m_{2,\min}\le m_{1,\min}$ by construction. The maximum primary mass is fixed to $m_{1,\max}=300\,M_\odot$ and is not varied in the inference.

\begin{table}[t]
\footnotesize
\begin{ruledtabular}
\begin{tabular}{lcl}
Parameter & Fiducial value & Prior \\
\midrule
$\alpha_1$        & $1.7$  & $\mathcal{U}(-4,12)$ \\
$\alpha_2$        & $4.5$  & $\mathcal{U}(-4,12)$ \\
$m_b$             & $36.0\,M_\odot$ & $\mathcal{U}(20,50)\,M_\odot$ \\
$\mu_1$           & $9.8\,M_\odot$  & $\mathcal{U}(5,20)\,M_\odot$ \\
$\sigma_1$        & $0.65\,M_\odot$ & $\mathcal{U}(0,10)\,M_\odot$ \\
$\mu_2$           & $33.0\,M_\odot$ & $\mathcal{U}(25,60)\,M_\odot$ \\
$\sigma_2$        & $3.9\,M_\odot$  & $\mathcal{U}(0,10)\,M_\odot$ \\
$m_{1,\min}$ & $2.23\,M_\odot$ & $\mathcal{U}(2,10)$ \\
$m_{2,\min}$ & $2.11\,M_\odot$  & $\mathcal{U}(2,m_{1,\min})$ \\
$\delta m_1$      & $4.3\,M_\odot$  & $\mathcal{U}(0,10)\,M_\odot$ \\
$\boldsymbol\lambda=(\lambda_0,\lambda_1,\lambda_2)$ & $(0.36,0.59,0.05)$ & $\mathrm{Dirichlet}(1,1,1)$ \\
$\beta$           & $1.2$  & $\mathcal{U}(-2,7)$ \\
$\delta m_2$      & $4.9\,M_\odot$  & $\mathcal{U}(0,10)\,M_\odot$ \\
\end{tabular}
\end{ruledtabular}
\caption{Fiducial values and priors for the primary- and
secondary-mass model parameters. 
$m_{1,\max}$ is
fixed and not sampled.}
\label{tab:mass_fiducial}
\end{table}

GW detectability and parameter estimation are modeled with the Fisher-matrix pipeline \textsc{GWFAST}~\cite{Iacovelli:2022bbs,Iacovelli:2022mbg}, using the \textsc{IMRPhenomHM} waveform~\cite{London:2017bcn}. We consider an A$+$-era LIGO--Virgo--KAGRA--LIGO India network~\cite{LIGOScientific2014pky,LIGOI,VIRGO2014yos,Aso2013eba}. The two US LIGO detectors and LIGO India use the \texttt{AplusDesign} sensitivity curve, Virgo uses \texttt{avirgo\_O5low\_NEW}, and KAGRA uses \texttt{kagra\_80Mpc}; the low-frequency cutoff is $10\,\mathrm{Hz}$. Sensitivity curves are available at Ref.~\cite{LVK_sensitivity}.

For each selected event, the single-event likelihood is approximated by a multivariate Gaussian with covariance given by the inverse Fisher matrix. We work in the zero-noise approximation. We draw $5000$ samples per event and impose the parameter-estimation priors used in the production pipeline, including priors flat in detector-frame component masses with $m_{2,D}<m_{1,D}$ and $\pi_{\rm PE}(d_L)\propto d_L^2$.

\subsection{Detection and localization selection}
\label{sec:gwselection}

The analysis deliberately targets the best-localized tail of the detected population. We require
\begin{equation}
\label{eqS:cuts}
\rho_{\rm net}>25,
\qquad
\Delta\Omega_{90}<2\,\mathrm{deg}^2,
\end{equation}
where $\rho_{\rm net}$ is the network matched-filter signal-to-noise ratio and $\Delta\Omega_{90}$ the $90\%$ credible sky area. The localization cut targets the regime in which explicit host sampling can produce concentrated galaxy posteriors while keeping the number of candidates computationally manageable. With the adopted setup, $N_{\rm gal}^i$ ranges from a few galaxies to $\mathcal O(10^4)$.

For each injected weighting model we retain $70$ events passing Eq.~\eqref{eqS:cuts}, corresponding approximately to the best-localized subset of $\sim200$ detections obtained over $\sim2\,$yr at $100\%$ duty cycle in the adopted idealized network. The $2\,\mathrm{deg}^2$ threshold is therefore a deliberate analysis choice rather than a claim about a typical O5 localization. 


Selection effects are evaluated using $10^7$ injections. The injection set covers the catalog sky area and extends to the network horizon. The same $\rho_{\rm net}$ and $\Delta\Omega_{90}$ thresholds in Eq.~\eqref{eqS:cuts} are applied to the injections; approximately $1.5\times10^5$ survive. The selection factor $\xi(\lambda)$ is then estimated by reweighted Monte Carlo integration over the surviving injections. This explicitly conditions the population inference on the stringent localization selection.

\section{Validation of population inference}
\label{secS:popvalidation}

This section collects the supporting tests for the population-level results shown in the Letter. The primary validation is recovery of the injected galaxy weighting, cosmology, and CBC population under the same selections used to define the observed sample consistently with the injection set.

\subsection{Recovery of $H_0$, $\epsilon$, and the CBC population}

\begin{figure}[t]
\centering
\includegraphics[width=0.45\textwidth]{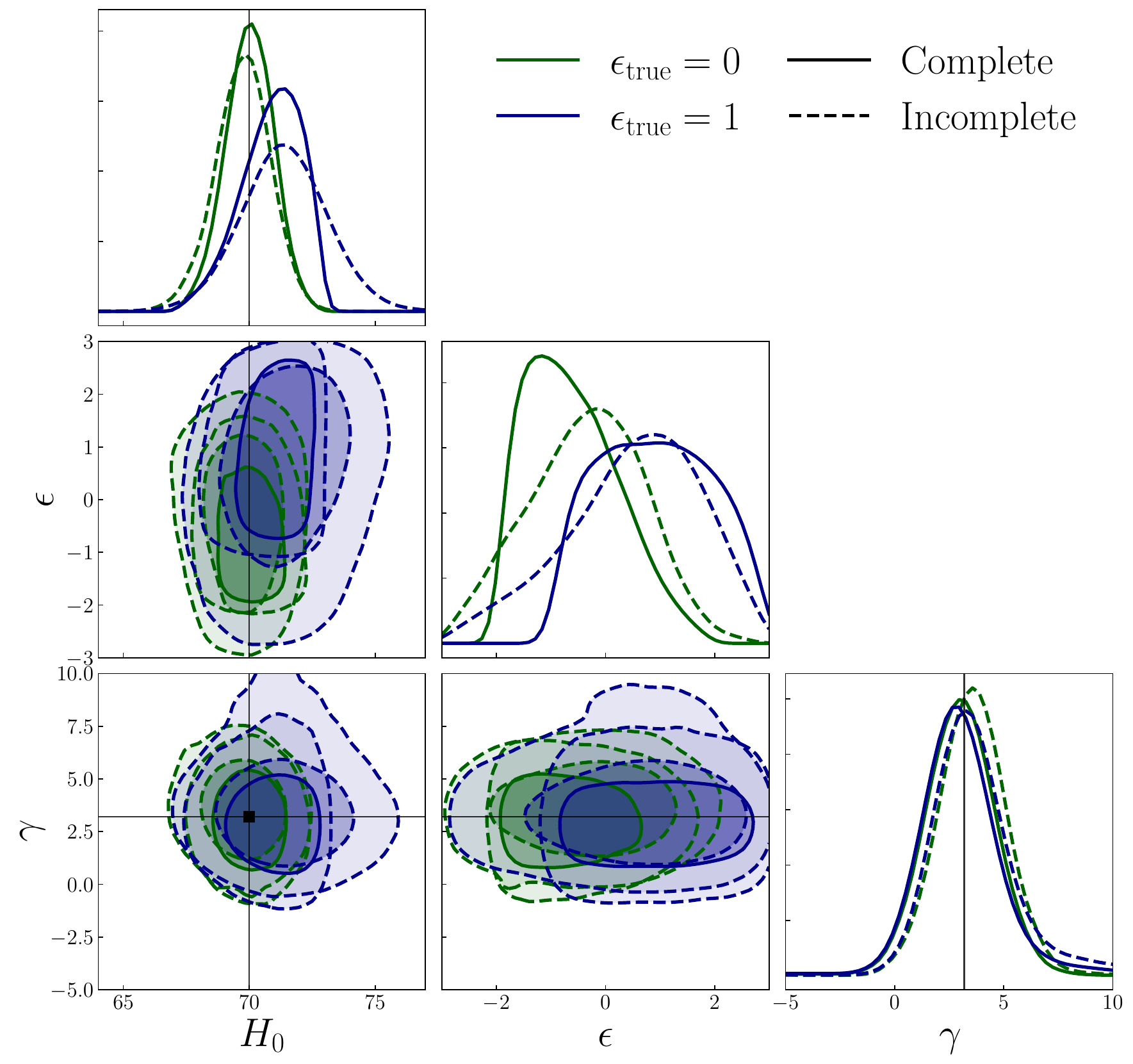}
\hfill
\includegraphics[width=0.45\textwidth]{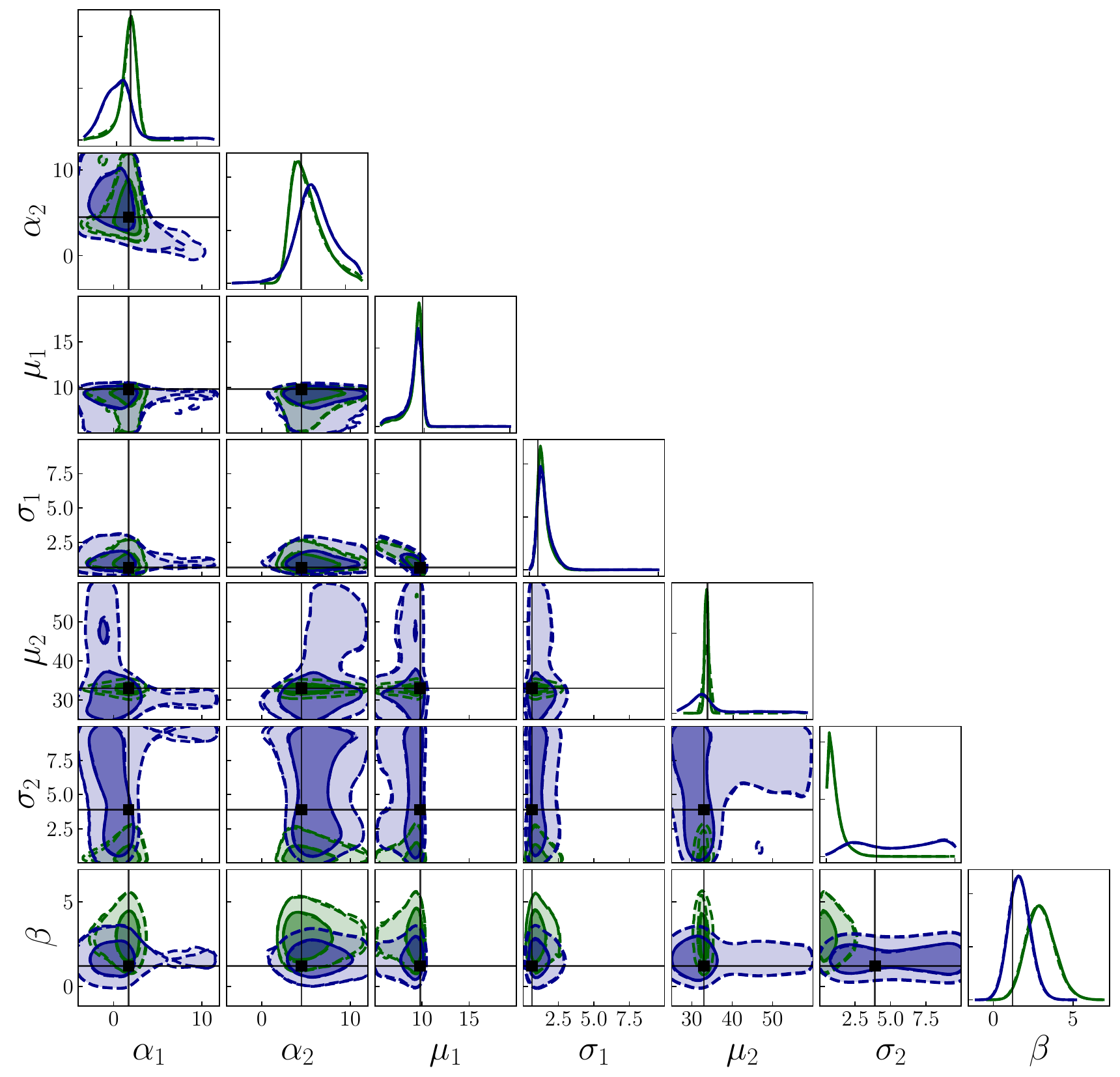}
\caption{Supporting population-level validation for the four simulated catalogs ($\epsilon_{\rm true}=0,1$; complete and incomplete galaxy catalogs). Joint and marginal posteriors for $H_0$, $\epsilon$, and $\gamma$ (left) and the main primary- and
secondary-mass model parameters (right), with 68\% and 95\% credible contours; injected values are marked by the black square and reference lines.
}
\label{figS:popfull}
\end{figure}

We repeat the full hierarchical analysis on four simulated catalogs, obtained by setting the fiducial galaxy-weighting parameter to $\epsilon_{\rm true}=0$ or $1$, and by using either complete or incomplete host-galaxy catalogs. Figure~\ref{figS:popfull} (left) shows the joint and marginal posteriors for $H_0$, $\epsilon$, and $\gamma$ for the four simulations, with the injected values marked by the black square and by vertical/horizontal lines~\footnote{Injected values for $\epsilon$ are reported in Figure~\ref{tabS:pop_recovery}.}. In all cases, the injected $H_0$, $\epsilon$, and $\gamma$ are recovered within the posterior support, and no statistically significant systematic offset between the complete and incomplete analyses is apparent at the resolution of this test. Table~\ref{tabS:pop_recovery} summarizes the posterior medians and 68\% credible intervals.

The injected $H_0=70\,\mathrm{km\,s^{-1}\,Mpc^{-1}}$ is recovered within the 68\% credible interval in all cases, with posterior medians ranging between $69.8$ and $71.4\,\mathrm{km\,s^{-1}\,Mpc^{-1}}$ across independent catalog realizations. The injected $\epsilon_{\rm true}$ is likewise recovered within the 68\% credible interval, although the posterior on $\epsilon$ remains broad. 
Finally, we do not find evidence for correlations between $H_0$, $\epsilon$, and $\gamma$ in the joint posteriors shown in Fig.~\ref{figS:popfull} (left).
\begin{table}[h]
\footnotesize
\begin{ruledtabular}
\begin{tabular}{lccc}
 & $H_0\;[\mathrm{km\,s^{-1}\,Mpc^{-1}}]$ & $\epsilon$ & $\gamma$ \\
\hline
$\epsilon_{\rm true}=0$, comp.   & $70.0^{+0.9}_{-1.0}$ & $-0.7^{+1.0}_{-0.8}$ & $3.0^{+1.5}_{-1.6}$ \\
$\epsilon_{\rm true}=0$, inc. & $69.8^{+1.0}_{-1.1}$ & $-0.3^{+1.0}_{-1.2}$ & $3.6^{+1.5}_{-1.5}$ \\
$\epsilon_{\rm true}=1$, comp.   & $71.0^{+1.1}_{-1.4}$ & $0.9^{+1.2}_{-1.1}$  & $2.9^{+1.7}_{-1.6}$ \\
$\epsilon_{\rm true}=1$, inc. & $71.4^{+1.7}_{-1.7}$ & $0.6^{+1.1}_{-1.4}$  & $3.4^{+1.9}_{-1.5}$ \\
\end{tabular}
\caption{Posterior recovery of $H_0$, $\epsilon$, and $\gamma$ for the simulated catalogs. Values quoted are posterior medians with 68\% credible intervals.}
\label{tabS:pop_recovery}
\end{ruledtabular}
\end{table}

As a complementary validation of the inference, we analyze the mass-model reconstruction (see Sec.~\ref{secS:sims}), comparing the posterior distributions of the primary- and secondary-mass parameters recovered in each of the four simulations against the injected fiducial values. Figure~\ref{figS:popfull} (right) shows that the recovered posteriors are compatible with the injected fiducial values across all four simulated populations within statistical uncertainties.



\section{Validation and characterization of host inference}
\label{sec:hostvalidation}

This section collects a set of diagnostics that assess and quantify the informativeness of the host inference. Together, they provide a comprehensive characterization of host-identification performance across our simulated populations. The main related results are shown in the main text.

First, given the full population of simulated GW events and the associated galaxy catalog, we study if the inferred ranking of candidates outperform random expectations trough a true-host rank statistics.
When the injected host is retained in the catalog, its host association performance can be quantified through its position in the inferred ranking. We define its integer rank as
\begin{equation}
\label{eqS:Rtrue}
R_{{\rm true},i}
=1+\sum_{g=1}^{N_{\rm gal}^i}
\mathbb I\!\left[p^{\rm cond}_{g,i}>p^{\rm cond}_{{\rm true},i}\right],
\end{equation}
i.e., one plus the number of catalog galaxies assigned a higher conditional probability than the true host, and the corresponding fractional rank as
\begin{equation}
\label{eqS:rtrue}
r_i\equiv\frac{R_{{\rm true},i}-1}{N_{\rm gal}^i},
\end{equation}
allowing comparison across events with different $N_{\rm gal}^i$. By construction, $R_{{\rm true},i}=1$ (equivalently $r_i=0$) indicates that the injected host is the highest-probability candidate in the catalog, while $R_{{\rm true},i}=N_{\rm gal}^i$ (equivalently $r_i\to1$) corresponds to it being ranked at the bottom.
%
We compare the cumulative distribution
\begin{equation}
F(<r)=P(r_i\leq r)\,,
\end{equation}
i.e. the fraction of such events whose true host falls within the top fraction $r$ of the ranked candidates,
with the expectation for a random ordering of the candidate galaxies. For event $i$, random ordering makes $R_{{\rm true},i}$ uniform on $\{1,\ldots,N_{\rm gal}^i\}$ and 
$F(<r)$ approaches the diagonal in the large-$N_{\rm gal}$ limit: $F_{\rm random}(<r)=r$. An observed cumulative distribution that lies systematically above this reference curve indicates that the true host is ranked higher than expected by chance, i.e. that the inference carries genuine host-identification power. 
%

This diagnostic is well defined only for the complete-catalog case, where the true host is known for every event and our framework can be validated in the regime where the amount of information to be recovered is maximal. Here, it is also particularly useful for probing the impact of fixed galaxy-weight prescriptions on host-association performance. Investigating the impact of different injected populations and catalog completeness on this statistics is left for future work. 

To further assess association performances, we report the probability of recovering the true host among the top-ranked candidates for a fixed number of candidates:
\begin{equation}
P(R_{\rm true}=1),
\qquad P(R_{\rm true}\leq10),
\end{equation}
and the fractions of events for which the true host lies in the top $1\%$ and $10\%$ of catalog candidates, for each of the four simulated populations.
For the incomplete-catalog cases, these fractions are computed only over the subset of events for which the true host is retained in the catalog. In every simulated scenario we substantially exceed the random-ordering expectation, confirming that the inferred ranking carries genuine host-identification information. 

%

Second, from the perspective of a single-event host posterior, we quantify its degree of concentration, i.e. the number of highest-ranked galaxies required to cover a fixed fraction of the total host posterior. This is directly relevant for targeted follow-up campaigns aiming to recover the true host with high confidence. We sort the absolute catalog probabilities as
\begin{equation}
p^{\rm abs}_{(1),i}\geq p^{\rm abs}_{(2),i}\geq\cdots
\end{equation}
and define
\begin{equation}
\label{eqS:Nalpha}
N_{\alpha,i}^{\rm cat}
\equiv
\min\left\{k:\sum_{j=1}^{k}p^{\rm abs}_{(j),i}\geq\alpha\right\},
\qquad
\alpha=0.5,0.9,0.95.
\end{equation}
We denote these by $N_{50}^{\rm cat}$, $N_{90}^{\rm cat}$, and $N_{95}^{\rm cat}$. In an incomplete catalog, the total probability available to catalog galaxies is $p_{{\rm in},i}$. Hence $N_{\alpha,i}^{\rm cat}$ is not achievable using catalog galaxies alone when $p_{{\rm in},i}<\alpha$; such events are recorded separately rather than assigning an artificial large $N_\alpha^{\rm cat}$.
These are the quantities used in the main text.

We summarize the main event-level host-identification results for the four benchmark configurations, obtained by combining the true-host ranking and posterior concentration diagnostics described above. The resulting performance metrics are reported in Tab.~\ref{tabS:hostsummary}. We stress that complete- and incomplete-catalog performances should not be directly compared, since they probe different regimes of candidate sparsity within GW localization region. Isolating the effect of catalog incompleteness itself is left to future work.

\begin{table}[t]
\footnotesize
\begin{ruledtabular}
\begin{tabular}{lcccccccc}
& $P(R_{\rm true}=1)\,[\%]$ & $P(R_{\rm true}\leq10)\,[\%]$ & top $1\%\,[\%]$ & top $10\%\,[\%]$ & $\bar N_{90}^{\rm cat}$ & $P(p_{\rm in}\geq0.9)\,[\%]$ & $\bar N_{\rm gal}$ \\
\hline
$\epsilon_{\rm true}=0$, comp. & 23 & 46 & 37 & 93  & 256 & 100 & 538 \\
$\epsilon_{\rm true}=0$, inc.  & 23 & 49 & 64 & 100 & 78  & 53  & 686 \\
$\epsilon_{\rm true}=1$, comp. & 16 & 54 & 39 & 90  & 232 & 100 & 574 \\
$\epsilon_{\rm true}=1$, inc.  & 19 & 56 & 60 & 98  & 117 & 51  & 714 \\
\end{tabular}
\caption{Summary of host-identification performance. For incomplete catalogs, rank statistics (columns 2--5) are evaluated only over events whose injected host is retained in the catalog, and $\bar N_{90}^{\rm cat}$ is evaluated only over events for which the corresponding probability level is achievable. $P(p_{\rm in}\geq0.9)$ denotes the fraction of events for which the posterior probability that the true host is represented in the catalog exceeds $90\%$; for complete catalogs this probability equals unity by construction. All median quantities across events are denoted by an overbar.}
\label{tabS:hostsummary}
\end{ruledtabular}
\end{table}


\section{GW170817 analysis and robustness}
\label{secS:gw170817}

In this section we validate the hierarchical framework on GW170817, using the event as a controlled test of its ability to infer host-galaxy properties and identify the most probable host without imposing the independently known electromagnetic counterpart. 


\subsection{GW data, galaxy catalog, and analysis choices}

We use the parameter-estimation samples of Ref.~\cite{Finstad:2018wid}, which retain the GW sky-localization information rather than fixing the source position to the electromagnetic counterpart. The known host NGC~4993 is therefore not imposed and can be used as an external validation of the inferred ranking.

The galaxy prior is constructed from GLADE+~\cite{Dalya:2021ewn} using $K$-band luminosities. Catalog redshifts are corrected for peculiar velocities; the reported peculiar-motion uncertainty is added in quadrature to the redshift uncertainty. Because the GW170817 sky localization is small, we neglect angular variation of the GLADE+ completeness and use $P_{\rm comp}(z,\epsilon)$.

Following the previous GW170817 dark-siren analysis of Ref.~\cite{LIGOScientific:2018gmd}, we consider two luminosity selections,
\begin{equation}
L>0.005L_*,\qquad L>0.36L_*,
\end{equation}
where $L_*$ is the characteristic $K$-band luminosity of the Schechter function used for the completeness calculation, whose parameters are taken from Ref.~\cite{Lu:2016vmu}. The lower threshold retains a much larger galaxy population but is less complete; the higher threshold gives a more complete sample relative to the selected luminosity range. The purpose of the comparison is to test robustness to the adopted luminosity cut, not to identify a preferred universal threshold. 

\begin{figure}[t]
\centering
\includegraphics[width=0.45\textwidth]{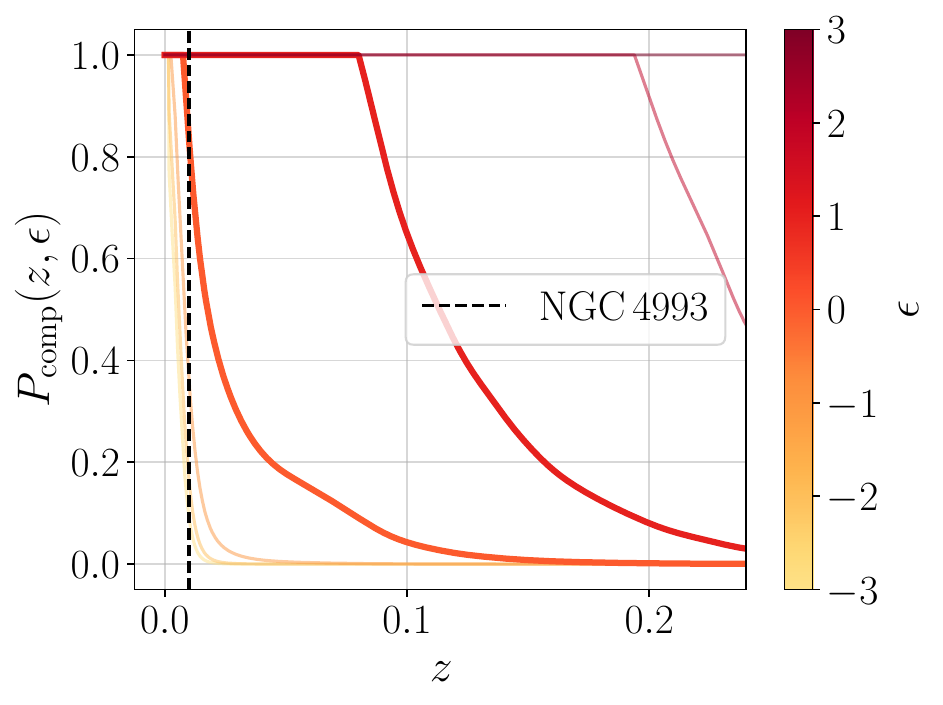}
\hfill
\includegraphics[width=0.45\textwidth]{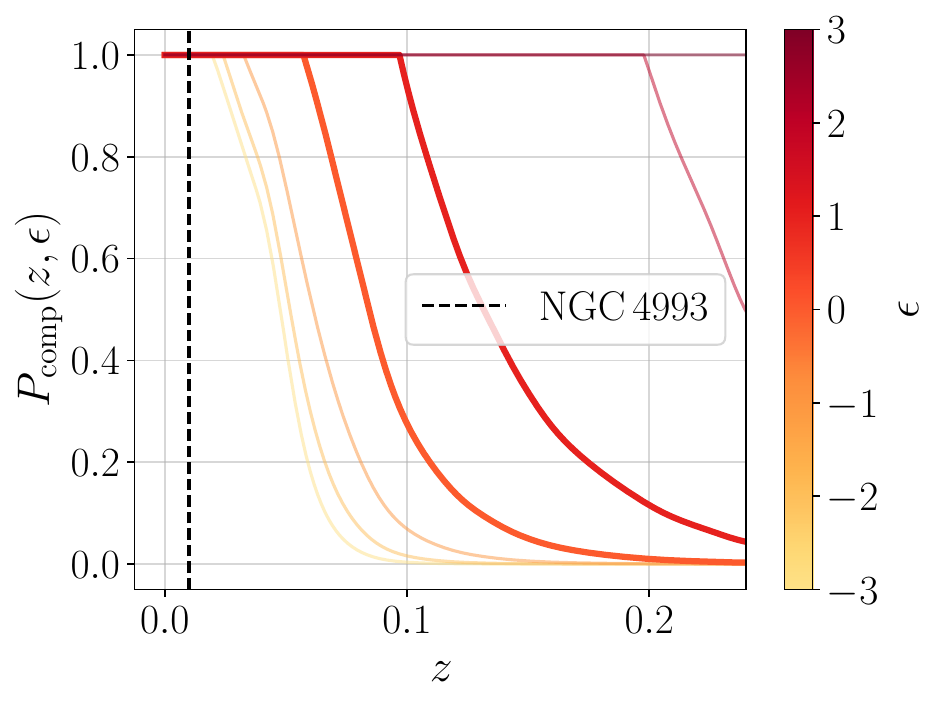}
\caption{$L > 0.005\,L_*$ (left) and $L > 0.36\,L_*$ (right) catalog completeness as function of redshift and luminosity-weight $\epsilon$. Thick lines represent completeness at $\epsilon=0$ and $\epsilon=1$. Dashed lines correspond to the redshift of the true host NGC 4993.}
\label{figS:gwcomp}
\end{figure}

For $L>0.005L_*$, the catalog completeness at $z\simeq0.01$, the redshift of NGC~4993, is approximately $83\%$ for $\epsilon=0$ and approximately $100\%$ for $\epsilon=1$. The catalog reaches $90\%$ incompleteness at $z\simeq0.08$ and $z\simeq0.19$, respectively. For $L>0.36L_*$, the catalog is approximately complete at $z\simeq0.01$ for both benchmark weightings and reaches $90\%$ incompleteness at $z\simeq0.13$ for $\epsilon=0$ and $z\simeq0.21$ for $\epsilon=1$ (see Figure~\ref{figS:gwcomp}).

Finally, in addition to these selections, we further explore an extreme $L>1.5L_*$ cut as a stress test in which the host prior is strongly dominated by the most luminous objects. The catalog is essentially fully complete at all redshifts considered and for both $\epsilon=0$ and 1.

We fix $\Omega_m=0.311$~\cite{Planck:2018vyg} and adopt flat priors
\begin{equation}
H_0\in[10,220]~\kmsMpc,\qquad\epsilon\in[-3,3].
\end{equation}
Candidate galaxies are selected inside the $99\%$ GW sky-localization region and inside a luminosity-distance interval extending $\pm3$ standard deviations around the GW distance estimate. The endpoints of the $H_0$ prior are used to convert this distance interval conservatively to redshift. This yields $69$ candidate hosts for $L>1.5L_*$, $182$  for $L>0.36L_*$, and $236$ for $L>0.005L_*$.

We marginalize over a Gaussian compact-binary mass distribution and over the Madau--Dickinson merger-rate parameters. We adopt uniform priors for the mass-distribution hyperparameters, $\mu_{m}\sim\mathcal{U}(1,2),M_\odot$ and $\sigma_{m}\sim\mathcal{U}(0.01,1),M_\odot$, and for the merger-rate parameters, $\gamma\sim\mathcal{U}(0,12)$, $z_p\sim\mathcal{U}(0,5)$, and $\kappa\sim\mathcal{U}(0,6)$. The GW selection factor is evaluated using the injection set employed in the corresponding dark-siren analysis~\cite{Mastrogiovanni:2023emh} available at \footnote{\href{https://zenodo.org/records/17093107}{zenodo.org/records/17093107}}.


\subsection{Robustness to the luminosity cut}

For the fiducial $L>0.005L_*$ selection, the analysis gives
\begin{equation}
\epsilon=1.1^{+0.9}_{-0.8},
\qquad
P(\epsilon>0)\simeq0.92,
\end{equation}
and
\begin{equation}
H_0=94.3^{+59.5}_{-32.6}~\kmsMpc,
\end{equation}
with quoted intervals corresponding to 68\% credible ranges consistent with previous dark-siren analyses~\citep{LIGOScientific:2018gmd,Mastrogiovanni:2023emh,Borghi:2023opd} and with the bright-siren constraint of~\cite{Abbott:2017xzu}. We find no strong correlation between $H_0$ and the luminosity-weighting parameter. The $H_0$ constraint is included here for completeness; the main text emphasizes the galaxy-weight and host-association results.

For the more restrictive $L>0.36L_*$ selection, the current analysis gives
\begin{equation}
\epsilon=1.2^{+0.9}_{-0.8},
\qquad
H_0=94.5^{+65.4}_{-32.9}~\kmsMpc,
\end{equation}
broadly consistent with the fiducial result. Increasing the luminosity threshold removes fainter galaxies and restricts the candidate population to more luminous objects, with a modest enhancement of the high-$H_0$ tail. The inferred $\epsilon$ should be interpreted relative to the galaxy population defined by the luminosity cut; changing that cut changes the population over which the host preference is parameterized, rather than constituting a purely technical catalog choice. 

\begin{figure}[t]
\centering
\includegraphics[width=.45\textwidth]{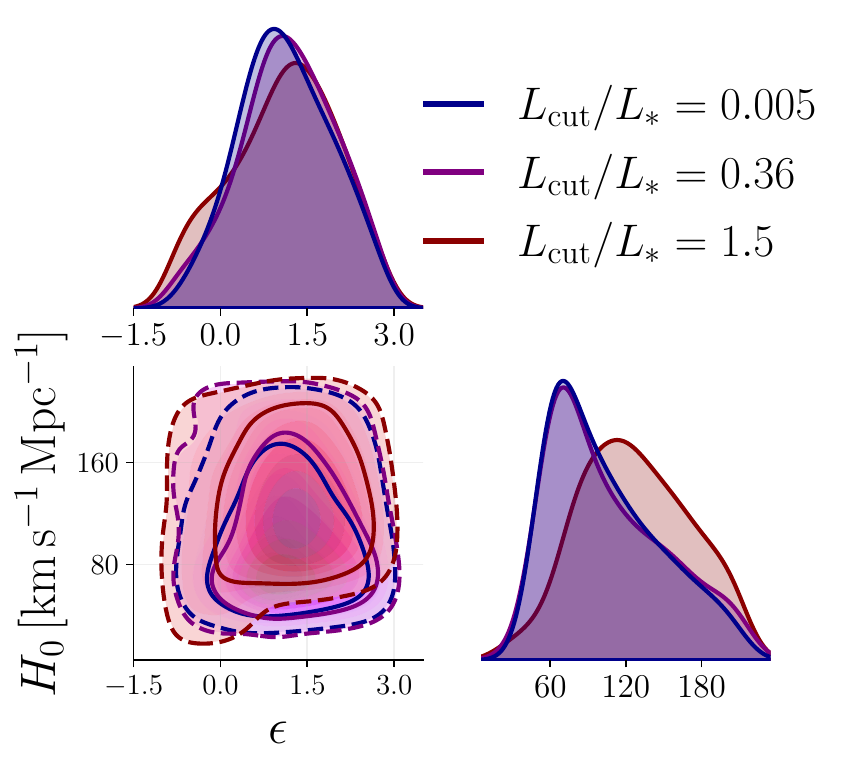}
\hfill
\includegraphics[width=.45\textwidth]{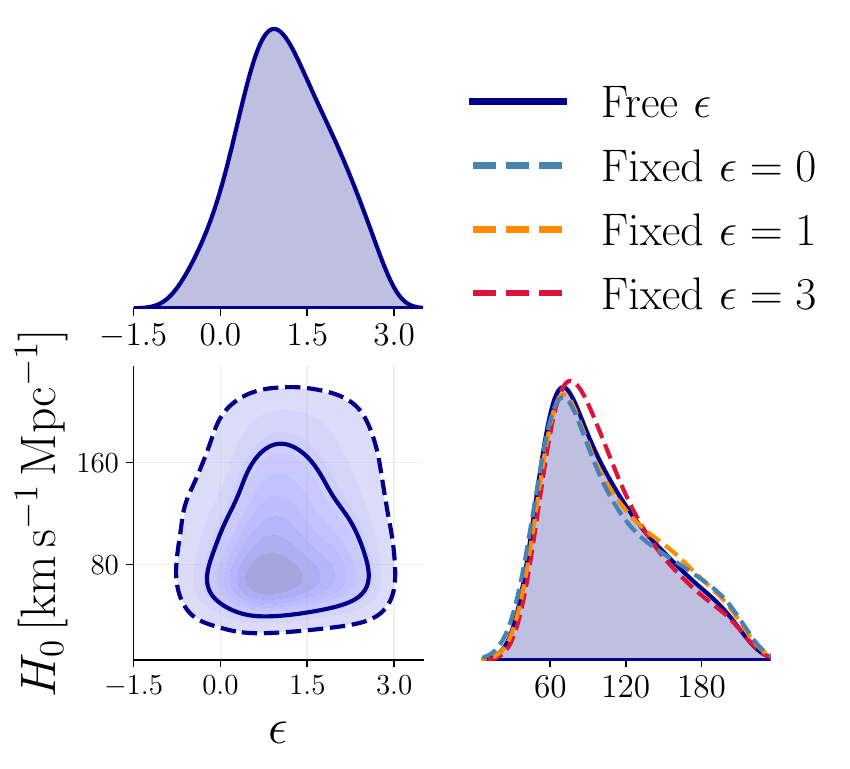}
\caption{GW170817 population-level inference and robustness to different luminosity thresholds and fixed luminosity weighting. Left: joint posterior for $\{H_0\,,\epsilon\}$ from GW170817 treated as a dark siren using GLADE+ galaxies with $K$-band luminosities comparing $L>0.005\,L_{*}$, $L>0.36\,L_{*}$, and $L>1.5\,L_{*}$ selection thresholds. Contours show the $68\%$ and $95\%$ credible regions. Right: posterior comparison within the $L>0.005\,L_{*}$ selection for free $\epsilon$, fixed $\epsilon=0$, $\epsilon=1$, and fixed $\epsilon=3$.}
\label{figS:gwcorner}
\end{figure}

For both selection choices, NGC~4993 is the highest-ranked catalog galaxy, with an absolute host probability of approximately $21\%$ for $L>0.005L_*$ and $29\%$ for $L>0.36L_*$. The posterior probability of the host being contained in the catalog decreases from approximately $98\%$ to $90\%$ when adopting the brighter luminosity threshold. 
The ranking is relatively concentrated in both cases, with the second- and third-ranked galaxies substantially less probable than NGC~4993. For the $L>0.005L_*$($L>0.36L_*$) selection, the cumulative probabilities reach $50\%$, $90\%$, and $95\%$ within $N_{50}=4(3)$, $N_{90}=26(14)$, and $N_{95}=42(24)$ galaxies, respectively, well below the total number of candidate galaxies $N_{\rm gal} = 236(182)$.

As an extreme stress test, we additionally investigate a substantially more restrictive threshold, $L>1.5L_*$, close to the luminosity of the true host. This cut retains only $69$ candidate galaxies, including the true one, and therefore probes a regime in which the host prior is strongly dominated by the most luminous objects. Despite this extreme cut, the population-level posteriors on $H_0$ and $\epsilon$ remain compatible with the other selections results, although the median $H_0$ is shifted upward by almost $30\,\mathrm{km\,s^{-1}\,Mpc^{-1}}$. While this shift occurs in a regime with large posterior uncertainties, it illustrates that overly aggressive luminosity cuts can potentially induce biases in the inferred cosmological parameters by substantially altering the host-galaxy association.

In this $L>1.5L_*$ case, the posterior becomes markedly more concentrated and the highest-ranked candidate is NGC~4548 with an absolute host probability of $\sim32\%$, while NGC~4993 is not among the highest-ranked candidates. The posterior becomes markedly more concentrated, with only $N_{90}=7$ galaxies.
The host inference results for the different luminosity-threshold selections are summarized in Table~\ref{tabS:host_summary}.

The change in the preferred host is particularly illustrative of the sensitivity to an aggressive luminosity selection.
Although NGC~4993 is intrinsically luminous, the $L>1.5L_*$ cut preferentially retains galaxies that are still brighter than the true host and substantially reduces the available host population. In the resulting sample, the most probable galaxy has $L/L_*\simeq2.5$ whereas the true host has $L/L_*\simeq1.7$. The inferred most-likely host therefore differs from the true host not because the latter is intrinsically faint, but because the aggressive luminosity threshold removes a large fraction of the candidate population and amplifies the relative weight of the most luminous remaining galaxies.

This case highlights an important limitation of interpreting luminosity cuts as a purely technical catalog choice. While the $L>0.005L_*$ and $L>0.36L_*$ selections yield broadly consistent host rankings and, crucially, both correctly identify NGC~4993 as the highest-ranked candidate, pushing the threshold to values close to the luminosity of the true host produces a qualitatively different inference, with the posterior becoming strongly concentrated on a small number of luminous galaxies and the highest-ranked candidate no longer corresponding to the true host. This contrast indicates that moderate luminosity cuts do not compromise host identification in this event, but that overly aggressive cuts carry a genuine risk of false host identification by artificially removing much of the candidate population and amplifying the relative weight of the most luminous galaxies.

\begin{table}[t]
\footnotesize
\begin{ruledtabular}
\begin{tabular}{lcccccccccc}
$L_{\rm cut}$ &
$p_{\rm cond}^{\rm NGC}$ [\%]&
$p_{\rm miss}$ [\%]&
$p_{\rm abs}^{\rm NGC}$ [\%]&
$\Delta p_2$ [\%]&
$\Delta p_3$ [\%]&
$N_{50}$ &
$N_{90}$ &
$N_{95}$ &
$N_{\rm gal}$ &
1-st ranked\\
\hline
$0.005L_*$ & $20.7$ & $1.9$ & $20.3$ & $\sim4$ & $\sim12$ & $4$ & $26$ & $42$ & 
$236$ & NGC~4993\\
$0.36L_*$  & $28.6$ & $10.0$ & $25.7$ & $\sim11$ & $\sim18$ & $3$ & $14$ & $24$ & 
$182$ & NGC~4993\\
$1.5L_*$  & $11.1$ & $26.0$ &$8.2$& -- & -- & $3$ & $7$ & $10$ & 
$69$ &  NGC~4548 \\
\end{tabular}
\caption{Summary of the host-galaxy ranking and posterior concentration for the two luminosity-threshold selections. $p_{\rm cond}^{\rm NGC}$ is the conditional host probability of NGC~4993, $p_{\rm miss}$ is the posterior probability that the host is not contained in the catalog, and $p_{\rm abs}^{\rm NGC}$ is the corresponding absolute host probability. $\Delta p_2$ and $\Delta p_3$ denote the absolute difference between the absolute probabilities of the second- and third-ranked galaxies and that of first-ranked one, NGC~4993. $N_{\rm gal}$ is the total number of possible candidate galaxies considered. $\Delta p_2$ and $\Delta p_3$ are not reported for the $L>1.5L_*$ case since NGC~4993 is not the first-ranked host.} 
\label{tabS:host_summary}
\end{ruledtabular}
\end{table}
\subsection{Sensitivity to the luminosity-weighting assumption}

The values $\epsilon=0$ and $\epsilon=1$ represent commonly adopted assumptions in the literature, and we therefore treat them as our baseline for a full quantitative comparison. The $H_0$ posteriors obtained by fixing $\epsilon=0$ and $\epsilon=1$ are shown in the $H_0$ marginal of Fig.~\ref{figS:gwcorner} as light-blue and red dotted curves, respectively. Relative to these two fixed-weighting cases, marginalizing over $\epsilon$ produces only a modest and statistically insignificant shift of the posterior's peak. 

On the other hand, NGC~4993 is not the highest-ranked catalog galaxy in both cases: it is the most probable catalog host for $\epsilon=1$, but not for $\epsilon=0$. In the latter case, the highest-ranked galaxy is ESO575-053, with an absolute host probability of $\sim 15\%$, while NGC~4993 is ranked 4-th with a probability of $\sim 9\%$. This change in ranking highlights the sensitivity of the inferred host preference to the assumed luminosity weighting; when no luminosity weighting is applied ($\epsilon=0$), the posterior probability is distributed more broadly among the catalog galaxies, and another candidate is favored over NGC~4993. Increasing the luminosity weighting to $\epsilon=1$ instead shifts probability toward NGC~4993, making it the highest-ranked catalog candidate.

The increase in the absolute probability of NGC~4993 is not solely due to the higher conditional probability assigned to this galaxy. It is also driven by the substantially smaller probability that the host lies outside the catalog, which decreases from $\sim 18\%$ for $\epsilon=0$ to zero for $\epsilon=1$. For $\epsilon=0$, a non-negligible fraction of the posterior probability is assigned to the possibility that the host is not present in the catalog, thereby reducing the absolute probability of all catalog galaxies, including NGC~4993. For $\epsilon=1$, instead, the missing-host component vanishes and the full posterior probability is contained within the catalog. The host posterior is also more concentrated for $\epsilon=1$. In particular, $N_{90}$ decreases from $29$ to $27$. 
These changes indicate that the effect of increasing the luminosity weighting is not limited to enhancing the probability of NGC~4993. Rather, probability is redistributed away from less-favored candidates, resulting in a more concentrated posterior over the catalog galaxies.

Going beyond state-of-the-art assumptions, we additionally consider a more extreme, non-standard value $\epsilon=3$. In this case, the luminosity prior completely dominates the inference: the true host, NGC~4993, is pushed to the second-ranked candidate ($p_{\rm abs} \sim 22\%$) in favor of a more luminous one, NGC~4830 ($p_{\rm abs} \sim 26\%$), which is almost 2 times more luminous than NGC~4993. Furthermore, the posterior becomes severely over-concentrated, with $N_{90}$ 
dropping to just 9.
As for the luminosity cut robustness test, the main host inference results for fixed luminosity-weighting within the fiducial selection are summarized in Table~\ref{tabS:host_summary_fix}. 

The progression from $\epsilon=0$ to $\epsilon=3$ perfectly illustrates an important caveat of imposing a fixed weighting. NGC~4993 has a relatively high luminosity, with $L/L_*\simeq1.7$, placing it among the top $\sim31\%$ most luminous galaxies in the candidate host distribution. In contrast, ESO575-053 has a much lower luminosity, with $L/L_*\simeq0.2$, placing it among the bottom $\sim10\%$ least luminous galaxies. Thus, for $\epsilon=0$ the most-likely galaxy is almost a factor of $8$ less luminous than the true host, while for $\epsilon=3$ a galaxy among the top $\sim12\%$ most luminous is instead favored. Increasing the luminosity weighting can therefore substantially enhance the posterior probability of intrinsically luminous galaxies and lead to false host identifications, with the identity of the highest-ranked galaxy changing across the different cases. Marginalizing over the luminosity-weighting parameter, rather than fixing it a priori, therefore provides a more conservative approach to dark-siren analyses, where the host-galaxy luminosity is itself used as a prior proxy for the probability of hosting the source.

\begin{table}[t]
\footnotesize
\begin{ruledtabular}
\begin{tabular}{lcccccccccccc}
&
$p_{\rm cond}^{\rm NGC}$ [\%]&
$p_{\rm miss}$ [\%]&
$p_{\rm abs}^{\rm NGC}$ [\%]&
$\Delta p_2$ [\%]&
$\Delta p_3$ [\%]&
$N_{50}$ &
$N_{90}$ &
$N_{95}$ &
$N_{\rm gal}$ & 1-st ranked\\
\hline
$\epsilon = 0$ & $8.7$ & $18.3$ & $7.1$ & -- & -- & $5$ & $29$ & $43$ & 
$236$ & ESO575-053\\
$\epsilon = 1$ & $23.6$ & $0$ & $23.6$ & $\sim10$ & $\sim14$ & $4$ & $27$ & $45$ & 
$236$ & NGC~4993 \\
$\epsilon = 3$ & $21.5$ & $0$ & $21.5$ & -- & -- & $3$ & $9$ & $16$ & 
$236$ & NGC~4830\\
\end{tabular}
\caption{Summary of the host-galaxy ranking and posterior concentration for fixed luminosity weightings. $\Delta p_2$ and $\Delta p_3$ are not reported for the $\epsilon=0$ and $\epsilon=3$ cases since NGC~4993 is not the first-ranked host.}
\label{tabS:host_summary_fix}
\end{ruledtabular}
\end{table}


\subsection{Candidate-set and implementation stability}

We additionally test the stability of the inference to the choice of the candidate set by moderately enlarging the sky and redshift region used to select potential host galaxies. This test is particularly relevant for $\epsilon$, since the normalization of the candidate-galaxy proposal depends on the luminosity weights. For GW170817, we repeat the inference with progressively enlarged candidate regions and assess the stability of $\epsilon$, $H_0$, and the leading host probabilities.

Candidate galaxies are selected in two steps. We first include galaxies within $\pm5$ standard deviations of the mean GW luminosity distance and within the $99\%$ GW sky-localization region, yielding $423$ candidate hosts for the $L>0.005,L_*$ selection. We then enlarge the sky region to the $5\sigma$ localization region, increasing the number of candidate galaxies to $795$.

Both configurations yield consistent posteriors for the headline parameters: $\epsilon = 1.1^{+0.9}_{-0.8}$ (423 candidates) and $\epsilon = 1.2^{+0.9}_{-0.8}$ (795 candidates), together with $H_0=95.6^{+61.5}_{-34.0}~\kmsMpc$ and $H_0=103.8^{+63.3}_{-42.9}~\kmsMpc$. The posterior shifts are non significant and well within sampling noise. In both cases the highest-probability host recovered by the inference is NGC 4993, with a comparable leading absolute probability that decreases mildly from $p^{\rm abs}_{\max}\simeq18\%$ to $p^{\rm abs}_{\max}\simeq14\%$ as the candidate set is enlarged
and the true host is diluted among a larger number of competing candidates.


\section{Numerical convergence and computational performance}
\label{secS:numerics}

Because the method samples both high-dimensional continuous parameters and discrete host states, we validate convergence at both levels exploiting complementary diagnostics and single-event level consistency checks. The discrete-indicator diagnostics is particularly important in order to verify a well-calibrated host inference an the reliability of probabilities associated to single-event candidate galaxies or missing hypoteses.


\subsection{Continuous parameters}
For continuous population, cosmological, and event-level parameters, the primary convergence criterion adopted is the rank-normalized split $\widehat{R}$~\cite{10.1214/ss/1177011136}. We require $\widehat{R}<1.01$ for all hyperparameters of scientific interest as a necessary condition for considering the chains converged. Since a satisfactory $\widehat{R}$ does not by itself rule out all Hamiltonian-sampling pathologies, we complement this primary criterion with the following additional checks:

\begin{itemize}
\item Effective Sample Size (ESS): as a practical production
criterion, we strictly require bulk and tail ESS $>100$ for every sampled
continuous hyperparameter, while aiming for ESS $\gtrsim 1000$ for the
headline parameters $H_0$ and $\epsilon$, as well as for the main CBC
population hyperparameters.
\item Monte Carlo Standard Error (MCSE): for the headline
parameters, the MCSE should be small compared with the posterior
uncertainty, with a target of less than roughly $5\%$ of the posterior
standard deviation (SD).
\end{itemize}

Table~\ref{tab:convergence} summarizes, for the final production runs
related to the main text results, the worst $\widehat{R}$, the
minimum/median ESS, and the MCSE/SD ratio.

\begin{table}[t]
\footnotesize
\begin{ruledtabular}
\begin{tabular}{lccccc}
Run & $\widehat R_{\max}$ & ESS$_{\rm bulk}^{\min}$ & ESS$_{\rm tail}^{\min}$ &
ESS$_{\rm bulk}^{\rm head,\min}$ & MCSE/SD$^{\rm head}_{\max}\,[\%]$ 
\\
\hline
$\epsilon_{\rm true}=0$, comp. & 1.005 & $\sim 2\times 10^3$ & $\sim 1\times 10^3$ & $\sim 2\times 10^3$ & 2 
\\
$\epsilon_{\rm true}=1$, comp. & 1.008 & $\sim 1\times 10^3$ & $\sim 3\times 10^3$ & $\sim 1\times 10^3$ & 4
\\
$\epsilon_{\rm true}=0$, incomp. & 1.009 & $\sim 6\times 10^2$ & $\sim 1\times 10^3$ & $\sim 6\times 10^2$ & 4 
\\
$\epsilon_{\rm true}=1$, incomp. & 1.009 & $\sim 5\times 10^2$ & $\sim 1\times 10^3$ & $\sim 5\times 10^2$ & 4 
\\
GW170817 & 1.002 & $\sim 1\times 10^3$  & $\sim 2\times 10^3$ & $\sim 1\times 10^3$ 
\\
\hline
Criteria & $<1.01$ & $>100$ & $>100$ & $\gtrsim 1000$ & $\lesssim5$ 
\\
\end{tabular}
\caption{Convergence and sampling-quality diagnostics across the final production runs in the main text. $\widehat R_{\max}$ and ESS values are computed over
all continuous hyper-parameters; ESS$_{\rm bulk}^{\rm head}$/ESS$_{\rm
tail}^{\rm head}$ and MCSE/SD refer specifically to the headline
parameters $H_0$, $\epsilon$, and the main CBC-population
hyperparameters. 
}
\label{tab:convergence}
\end{ruledtabular}
\end{table}

\subsection{Discrete host variables}

For discrete host variables, we complement the continuous-parameter diagnostics with categorical diagnostics based on the posterior probabilities assigned to individual host states. The integer value of $g_i$ has no meaningful metric ordering, so $\widehat R$ is not computed on the label itself. For candidate galaxies with non-negligible posterior support we instead define the binary chain
\begin{equation}
I^{(c)}_{g,i}=\mathbb I[g_i^{(c)}=g]\,,
\end{equation}
and compute $\widehat R$ and ESS for these indicators. For incomplete catalogs we apply the same test to $\mathbb I[s_i^{(c)}=0]$. These diagnostics directly test whether independent chains agree on the posterior mass assigned to the relevant physical host states.
We used to define ``non-negligible'' as the highest-ranked galaxies carrying the $90\%$ of the cumulative host probability provided their binary traces are sampled often enough to give a meaningful diagnostic. A practical requirement is an expected
$\sim 100$ visits per chain, i.e. $N_{\rm post}^{(c)}\,p_{g,i}^{(c)}\gtrsim100$, where $N_{\rm post}^{(c)}$ is the
number of post-warmup draws per chain. 

The $\widehat R$ and ESS values reported in Table~\ref{tab:convergence} include these discrete-indicator diagnostics alongside the continuous hyperparameters, confirming consistent convergence across both parameter types.

\putbib[references,\jobname Notes]

\end{bibunit}

\end{document}